\pdfoutput=1
\documentclass[12pt,a4paper]{article}

\usepackage{ifthen} 
\newboolean{pdflatex}
\setboolean{pdflatex}{true} 

\newboolean{articletitles}
\setboolean{articletitles}{true} 

\newboolean{uprightparticles}
\setboolean{uprightparticles}{false} 

\def\paperauthors{LHCb collaboration} 
\def\paperasciititle{Amplitude analysis of the B0->Kstar0 mu mu decay} 
\def\papertitle{Amplitude analysis of the $B^{0}\to K^{*0}\mu^+\mu^-$ decay} 
\def\paperkeywords{{High Energy Physics}, {LHCb}} 
\def\papercopyright{\the\year\ CERN for the benefit of the LHCb collaboration} 
\def\paperlicence{CC BY 4.0 licence}
\def\paperlicenceurl{https://creativecommons.org/licenses/by/4.0/}

\def\thetak {\ensuremath{\theta_{K}}\xspace}

\usepackage[top=1in, bottom=1.25in, left=1in, right=1in]{geometry}

\usepackage{microtype}
\usepackage{lineno}  
\usepackage{xspace} 
\usepackage{caption} 

\usepackage{graphicx}  
\usepackage{color}
\usepackage{colortbl}
\graphicspath{{./figs/}} 

\usepackage{amsmath} 
\usepackage{amssymb}
\usepackage{amsfonts}
\usepackage{upgreek} 

\newcommand*\patchAmsMathEnvironmentForLineno[1]{%
\expandafter\let\csname old#1\expandafter\endcsname\csname #1\endcsname
\expandafter\let\csname oldend#1\expandafter\endcsname\csname
end#1\endcsname
 \renewenvironment{#1}%
   {\linenomath\csname old#1\endcsname}%
   {\csname oldend#1\endcsname\endlinenomath}%
}
\newcommand*\patchBothAmsMathEnvironmentsForLineno[1]{%
  \patchAmsMathEnvironmentForLineno{#1}%
  \patchAmsMathEnvironmentForLineno{#1*}%
}
\AtBeginDocument{%
\patchBothAmsMathEnvironmentsForLineno{equation}%
\patchBothAmsMathEnvironmentsForLineno{align}%
\patchBothAmsMathEnvironmentsForLineno{flalign}%
\patchBothAmsMathEnvironmentsForLineno{alignat}%
\patchBothAmsMathEnvironmentsForLineno{gather}%
\patchBothAmsMathEnvironmentsForLineno{multline}%
\patchBothAmsMathEnvironmentsForLineno{eqnarray}%
}

\usepackage{hyperxmp}

\usepackage[pdftex,
            pdfauthor={\paperauthors},
            pdftitle={\paperasciititle},
            pdfkeywords={\paperkeywords},
            pdfcopyright={Copyright (C) \papercopyright},
            pdflicenseurl={\paperlicenceurl}]{hyperref}

\usepackage[colorinlistoftodos,textsize=scriptsize]{todonotes}

\usepackage[bottom,flushmargin,hang,multiple]{footmisc}

\usepackage[all]{hypcap} 

\usepackage{xspace} 
\usepackage{upgreek}

\def\lhcb   {\mbox{LHCb}\xspace}

\def\MagUp {\mbox{\em Mag\kern -0.05em Up}\xspace}

\ifthenelse{\boolean{uprightparticles}}%
{

 \def\Pmu         {\ensuremath{\upmu}\xspace}

 \def\Ppsi        {\ensuremath{\uppsi}\xspace}

 \def\PDelta      {\ensuremath{\Delta}\xspace}                 
 \def\PXi         {\ensuremath{\Xi}\xspace}                 
 \def\PLambda     {\ensuremath{\Lambda}\xspace}                 
 \def\PSigma      {\ensuremath{\Sigma}\xspace}                 
 \def\POmega      {\ensuremath{\Omega}\xspace}                 
 \def\PUpsilon    {\ensuremath{\Upsilon}\xspace}
 \let\oldPi\Pi
 \def\PPi         {\ensuremath{\oldPi}\xspace}

 \def\PB      {\ensuremath{\mathrm{B}}\xspace}                 
                  
 \def\PD      {\ensuremath{\mathrm{D}}\xspace}

 \def\PJ      {\ensuremath{\mathrm{J}}\xspace}                 
 \def\PK      {\ensuremath{\mathrm{K}}\xspace}

 \def\Pb      {\ensuremath{\mathrm{b}}\xspace}

 \def\Pi      {\ensuremath{\mathrm{i}}\xspace}

 \def\Ps      {\ensuremath{\mathrm{s}}\xspace}

 \def\thebaroffset{0.0em}
}
{

 \def\Pmu         {\ensuremath{\mu}\xspace}

 \def\Ppsi        {\ensuremath{\psi}\xspace}                 
                  
 \mathchardef\PDelta="7101
 \mathchardef\PXi="7104
 \mathchardef\PLambda="7103
 \mathchardef\PSigma="7106
 \mathchardef\POmega="710A
 \mathchardef\PUpsilon="7107
 \mathchardef\PPi="7105
                  
 \def\PB      {\ensuremath{B}\xspace}                 
                  
 \def\PD      {\ensuremath{D}\xspace}

 \def\PJ      {\ensuremath{J}\xspace}                 
 \def\PK      {\ensuremath{K}\xspace}

 \def\Pb      {\ensuremath{b}\xspace}

 \def\Pi      {\ensuremath{i}\xspace}

 \def\Ps      {\ensuremath{s}\xspace}

 \def\thebaroffset{0.18em}
}
\newcommand{\offsetoverline}[2][\thebaroffset]{\kern #1\overline{\kern -#1 #2}}%

\makeatletter
\ifcase \@ptsize \relax
  \newcommand{\miniscule}{\@setfontsize\miniscule{4}{5}}
\or
  \newcommand{\miniscule}{\@setfontsize\miniscule{5}{6}}
\or
  \newcommand{\miniscule}{\@setfontsize\miniscule{5}{6}}
\fi
\makeatother

\DeclareRobustCommand{\optbar}[1]{\shortstack{{\miniscule (\rule[.5ex]{1.25em}{.18mm})}
  \\ [-.7ex] $#1$}}

\def\mup        {{\ensuremath{\Pmu^+}}\xspace}
\def\mun        {{\ensuremath{\Pmu^-}}\xspace} 

\def\mumu       {{\ensuremath{\Pmu^+\Pmu^-}}\xspace}

\def\squark    {{\ensuremath{\Ps}}\xspace}

\def\bquark    {{\ensuremath{\Pb}}\xspace}

\def\kaon    {{\ensuremath{\PK}}\xspace}

\def\KorKbar {\kern \thebaroffset\optbar{\kern -\thebaroffset \PK}{}\xspace}

\def\Kstarz  {{\ensuremath{\kaon^{*0}}}\xspace}

\def\Kstar   {{\ensuremath{\kaon^*}}\xspace}

\def\D       {{\ensuremath{\PD}}\xspace}

\def\DorDbar {\kern \thebaroffset\optbar{\kern -\thebaroffset \PD}\xspace}

\def\Dp      {{\ensuremath{\D^+}}\xspace}
\def\Dm      {{\ensuremath{\D^-}}\xspace}

\def\DpDm    {\ensuremath{\Dp {\kern -0.16em \Dm}}\xspace}

\def\B       {{\ensuremath{\PB}}\xspace}
\def\Bbar    {{\ensuremath{\offsetoverline{\PB}}}\xspace}

\def\BorBbar {\kern \thebaroffset\optbar{\kern -\thebaroffset \PB}\xspace}
\def\Bz      {{\ensuremath{\B^0}}\xspace}
\def\Bzb     {{\ensuremath{\Bbar{}^0}}\xspace}
\def\Bd      {{\ensuremath{\B^0}}\xspace}

\def\BdorBdbar {\kern \thebaroffset\optbar{\kern -\thebaroffset \Bd}\xspace}
\def\Bu      {{\ensuremath{\B^+}}\xspace}

\def\Bp      {{\ensuremath{\Bu}}\xspace}

\def\Bs      {{\ensuremath{\B^0_\squark}}\xspace}

\def\BsorBsbar {\kern \thebaroffset\optbar{\kern -\thebaroffset \Bs}\xspace}

\def\jpsi     {{\ensuremath{{\PJ\mskip -3mu/\mskip -2mu\Ppsi}}}\xspace}
\def\psitwos  {{\ensuremath{\Ppsi{(2S)}}}\xspace}

\def\Y#1S{\ensuremath{\PUpsilon{(#1S)}}\xspace}

\def\LorLbar     {\kern \thebaroffset\optbar{\kern -\thebaroffset \PLambda}\xspace}

\def\to                 {\ensuremath{\rightarrow}\xspace}

\def\AT#1     {\ensuremath{A_{\mathrm{T}}^{#1}}\xspace}           

\def\C#1      {\ensuremath{\mathcal{C}_{#1}}\xspace}                       
\def\Cp#1     {\ensuremath{\mathcal{C}_{#1}^{'}}\xspace}                    
\def\Ceff#1   {\ensuremath{\mathcal{C}_{#1}^{\mathrm{(eff)}}}\xspace}        
\def\Cpeff#1  {\ensuremath{\mathcal{C}_{#1}^{'\mathrm{(eff)}}}\xspace}       
\def\Ope#1    {\ensuremath{\mathcal{O}_{#1}}\xspace}                       
\def\Opep#1   {\ensuremath{\mathcal{O}_{#1}^{'}}\xspace}                    

\newcommand{\aunit}[1]{\ensuremath{\text{\,#1}}}       

\newcommand{\tev}{\aunit{Te\kern -0.1em V}\xspace}
\newcommand{\gev}{\aunit{Ge\kern -0.1em V}\xspace}
\newcommand{\mev}{\aunit{Me\kern -0.1em V}\xspace}
\newcommand{\kev}{\aunit{ke\kern -0.1em V}\xspace}
\newcommand{\ev}{\aunit{e\kern -0.1em V}\xspace}
 
\newcommand{\mevc}{\ensuremath{\aunit{Me\kern -0.1em V\!/}c}\xspace}
\newcommand{\gevc}{\ensuremath{\aunit{Ge\kern -0.1em V\!/}c}\xspace}
\newcommand{\mevcc}{\ensuremath{\aunit{Me\kern -0.1em V\!/}c^2}\xspace}
\newcommand{\gevcc}{\ensuremath{\aunit{Ge\kern -0.1em V\!/}c^2}\xspace}
\newcommand{\gevgevcccc}{\ensuremath{\gev^2\!/c^4}\xspace} 

\def\fb   {\ensuremath{\aunit{fb}}\xspace}
\def\invfb   {\ensuremath{\fb^{-1}}\xspace}

\def\gsim{{~\raise.15em\hbox{$>$}\kern-.85em
          \lower.35em\hbox{$\sim$}~}\xspace}
\def\lsim{{~\raise.15em\hbox{$<$}\kern-.85em
          \lower.35em\hbox{$\sim$}~}\xspace}

\def\tell1  {TELL1\xspace}
\def\ukl1   {UKL1\xspace}

\newcommand{\lhcborcid}[1]{\href{https://orcid.org/#1}{\hspace*{0.1em}\raisebox{-0.45ex}{\includegraphics[width=1em]{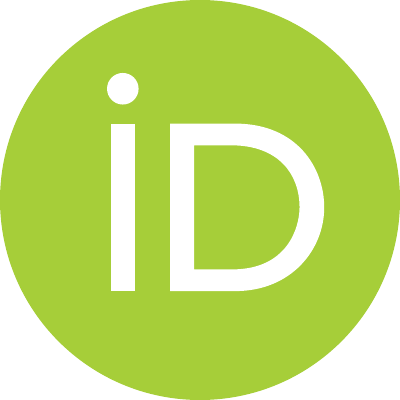}}}}

\usepackage{cite} 
\usepackage{mciteplus}

\usepackage{longtable} 

\usepackage{multirow}
\usepackage{makecell}
\usepackage{booktabs}
\usepackage{comment}

\begin{document}

\renewcommand{\thefootnote}{\fnsymbol{footnote}}
\setcounter{footnote}{1}


\begin{titlepage}
\pagenumbering{roman}

\vspace*{-1.5cm}
\centerline{\large EUROPEAN ORGANIZATION FOR NUCLEAR RESEARCH (CERN)}
\vspace*{1.5cm}
\noindent
\begin{tabular*}{\linewidth}{lc@{\extracolsep{\fill}}r@{\extracolsep{0pt}}}
\ifthenelse{\boolean{pdflatex}}
{\vspace*{-1.5cm}\mbox{\!\!\!\includegraphics[width=.14\textwidth]{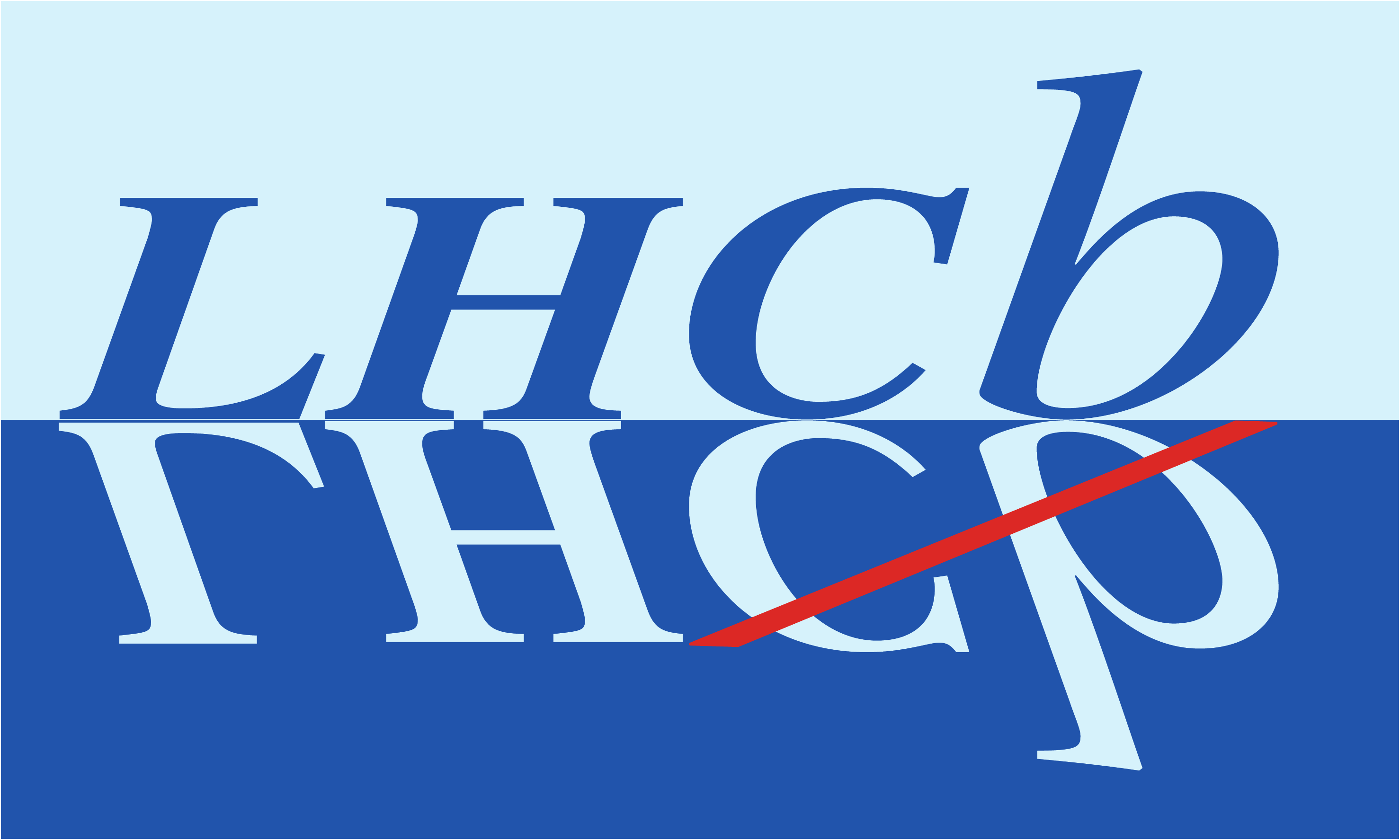}} & &}%
{\vspace*{-1.2cm}\mbox{\!\!\!\includegraphics[width=.12\textwidth]{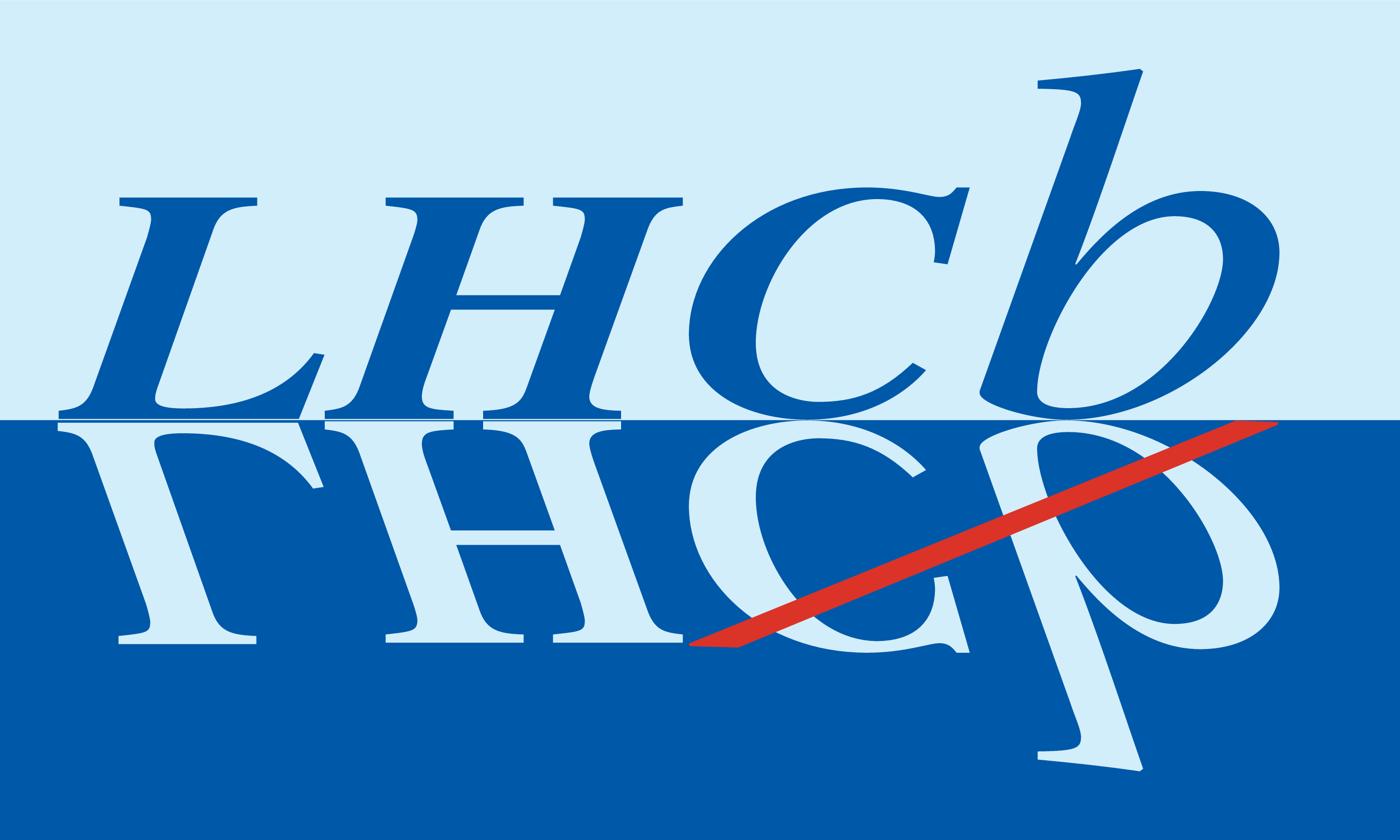}} & &}%
\\
 & & CERN-EP-2023-273 \\  
 & & LHCb-PAPER-2023-033 \\  
 & & \today \\ 
 & & \\
\end{tabular*}

\vspace*{4.0cm}

{\normalfont\bfseries\boldmath\huge
\begin{center}
  \papertitle 
\end{center}
}

\vspace*{2.0cm}

\begin{center}
\paperauthors\footnote{Authors are listed at the end of this Letter.}
\end{center}

\vspace{\fill}

\begin{abstract}
  \noindent
An amplitude analysis of the $B^{0}\to K^{*0}\mu^+\mu^-$ decay is presented using a
dataset corresponding to an integrated luminosity of $4.7\invfb$ of $pp$ collision data collected with the LHCb experiment. 
For the first time, the coefficients associated to short-distance physics effects,  sensitive to processes beyond the Standard Model, are extracted directly from the data through a $q^2$-unbinned amplitude analysis, where $q^2$ is the $\mu^+\mu^-$  invariant mass squared.
Long-distance contributions, which originate from non-factorisable QCD processes, are systematically investigated and the most accurate assessment to date of their impact on the physical observables is obtained. 
The pattern of measured corrections to the short-distance couplings is found to be consistent with previous analyses of \bquark- to \squark-quark transitions, with the largest discrepancy from the Standard Model predictions found to be at the level of 1.8 standard deviations.
The global significance of the observed differences in the decay is 1.4 standard deviations.

\end{abstract}

\vspace*{2.0cm}

\begin{center}
Published in Phys. Rev. Lett. 132 (2024) 131801
\end{center}

\vspace{\fill}

{\footnotesize 
\centerline{\copyright~\papercopyright. \href{\paperlicenceurl}{\paperlicence}.}}
\vspace*{2mm}

\end{titlepage}


\newpage
\setcounter{page}{2}
\mbox{~}
%
%
%
%


\renewcommand{\thefootnote}{\arabic{footnote}}
\setcounter{footnote}{0}

\cleardoublepage


\pagestyle{plain} 
\setcounter{page}{1}
\pagenumbering{arabic}



Flavour-changing neutral current (FCNC) processes  
involving the transition of a beauty into a strange quark provide powerful indirect probes for effects beyond the Standard Model (SM).
These transitions are mediated through virtual quantum loops in the SM, and as yet undiscovered particles may contribute to the transition amplitudes of the decay at a level comparable to  SM processes. These new contributions can cause deviations in the decay rate or in the angular distributions of the final-state particles.
Past studies of these processes have shown an intriguing set of discrepancies with respect to the SM predictions~\cite{LHCb-PAPER-2014-006,LHCb-PAPER-2013-017,LHCb-PAPER-2015-009,LHCb-PAPER-2022-050,LHCb-PAPER-2020-041}.
In particular, analyses of the $\Bz \to \Kstarz \mu^+\mu^-$ decay have 
reported statistically significant deviations in the branching fraction~\cite{LHCb-PAPER-2016-012} and in the  angular distributions of the decay~\cite{LHCB-PAPER-2013-019,LHCB-PAPER-2013-037,LHCB-PAPER-2015-051,LHCB-PAPER-2020-002,Belle:2016fev,Aaboud:2018krd,Sirunyan:2017dhj},
notably in observables with reduced theoretical uncertainties, \textit{e.g.}
in the so-called $P^{\prime}_{5}$ observable~\cite{Descotes-Genon:2015uva}.

\begin{figure}[t]
\centering
\includegraphics[width=0.65\textwidth]{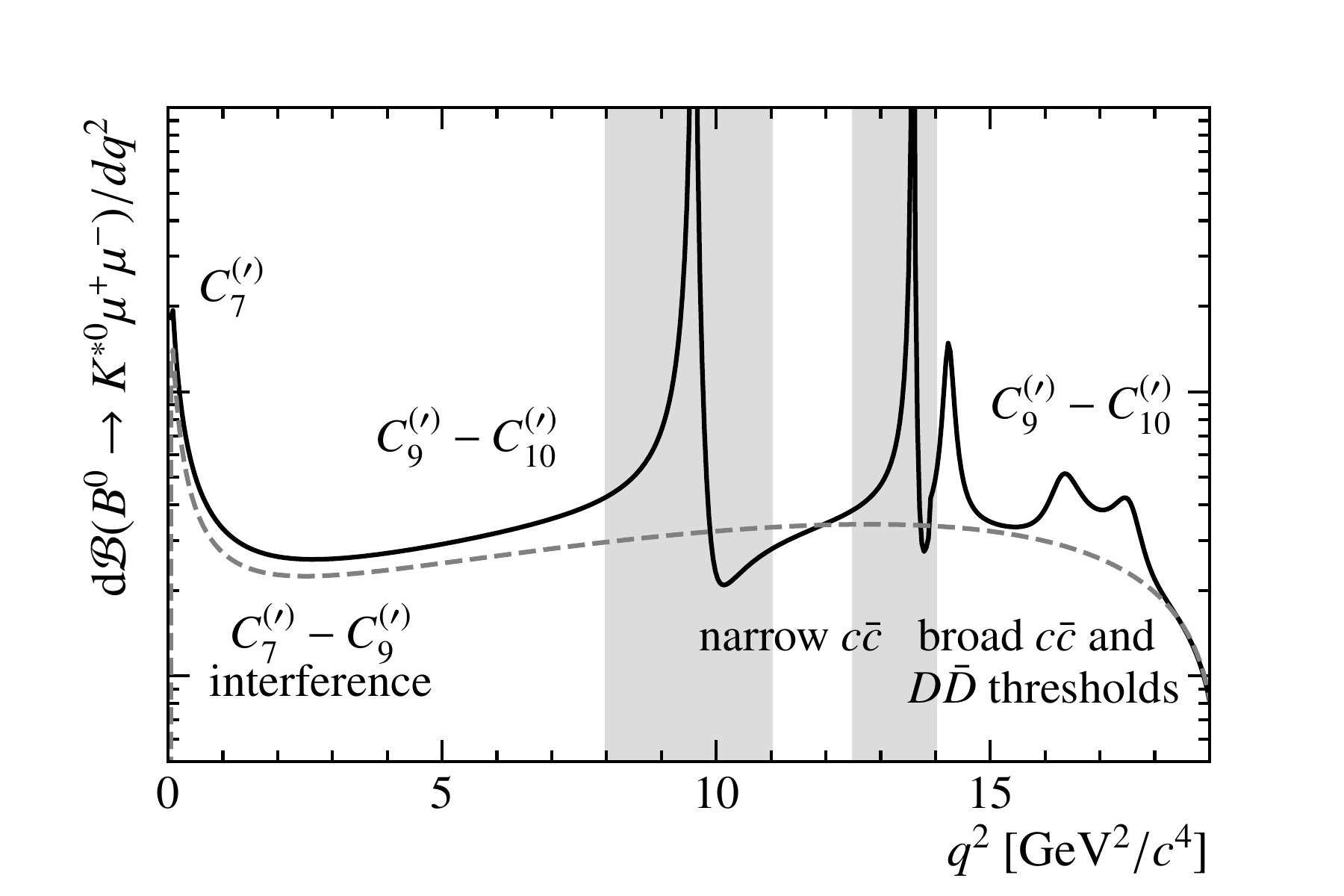}
\caption{Illustration of the $q^2$ spectrum in $B^0\to K^{*0}\mumu$ decays. 
The dashed line corresponds to the pure rare semileptonic decay, while the solid line includes the impact of different charmonium resonances. 
The gray bands correspond to the regions dominated by $B^0\to \jpsi K^{*0}$ and $B^0\to \psitwos K^{*0}$ tree-level decays. 
Magnitudes and phases of $c\bar{c}$ resonant components have been arbitrarily chosen for illustrative purposes. 
The dominant Wilson coefficients in each region of the spectrum are also highlighted for reference.
 \label{fig:q2}}
\end{figure}

Beauty- to strange-quark FCNC processes are typically described within an effective field theory~\cite{Buchalla:1995vs} in terms of four-fermion interactions.
Within this framework, the strengths of the different types of interaction are encoded by the so-called Wilson coefficients, labelled as $\mathcal{C}^{(\prime)}_i$, and beyond the Standard Model (BSM) effects would appear as a deviation in the values of one or more Wilson coefficients.
The most relevant contributions are from the so-called electromagnetic dipole operator and from vector and axial-vector interactions as shown in Fig.~\ref{fig:q2}, with coefficients ${\mathcal{C}}_{7}$, ${\mathcal{C}}_{9}$ and ${\mathcal{C}}_{10}$, respectively, and their counterparts with the opposite chirality, ${\mathcal{C}}_i^\prime$, which are suppressed in the SM. 
Global analyses of $b\to s$ transitions
have reported deviations from the SM expectations with significances as large as 4 standard deviations~\cite{Gubernari:2022hxn,Greljo:2022jac,Alguero:2023jeh,Ciuchini:2022wbq}.  
The largest tensions are seen in the real part of $\mathcal{C}_9$.  
The interpretation of the global analyses is complicated by large theoretical uncertainties on the prediction of non-factorisable (long-distance) hadronic contributions that  can mimic BSM effects~\cite{Jager:2012uw,Jager:2014rwa}.
In order to improve the sensitivity of the global analyses, it is imperative to find reliable ways of reducing the impact of these uncertainties.
Since the first measurement of $P^{\prime}_{5}$ by LHCb in 2013~\cite{LHCB-PAPER-2013-037}, 
progress has been made on the uncertainties of SM predictions~\cite{Khodjamirian:2017fxg,Straub:2015ica,Agadjanov:2016fbd,Gubernari:2018wyi,Descotes-Genon:2019bud,Gao:2019lta,Bouchard:2013eph,Bailey:2015dka,Horgan:2013hoa,Horgan:2015vla,Gubernari:2023puw,Khodjamirian:2010vf,Khodjamirian:2012rm,Descotes-Genon:2023ukb,Gubernari:2020eft,Bobeth:2017vxj}, 
and the most recent developments suggest that it is now possible to control the size of the long-distance contributions in data~\cite{Gubernari:2022hxn}. 
Nevertheless, no definitive consensus on the characterisation of these effects has been reached yet~\cite{Ciuchini:2022wbq}.

This Letter reports the first unbinned amplitude analysis of the decay $\Bz \to \Kstarz \mu^+\mu^-$ that determines simultaneously the short- and long-distance contributions.\footnote{Throughout this Letter, $\Kstarz$ is used to refer to
the $\Kstar(892)^{0}$ resonance and the inclusion of charge-conjugate processes is implied.}
The strategy of this analysis is based on the methodology discussed in Refs.~\cite{Bobeth:2017vxj,Gubernari:2022hxn,Chrzaszcz:2018yza,Hurth:2017sqw},
and allows a systematic evaluation of the impact of theoretical uncertainties on the observables of interest.
A more comprehensive description of these studies is reported in a companion article~\cite{LHCb-PAPER-2023-032}.

The differential decay rate for the $\Bz \to \Kstarz  \mu^+\mu^-$
decay, where the \Kstarz meson is reconstructed through the decay $\Kstarz  \to K^+ \pi^-$, is
fully described by five kinematic variables:  
the invariant mass squared of the $\mumu$ system, $q^2$, 
the invariant mass squared of the $K^+\pi^-$ system, $k^2$, 
and the three angles 
$\vec{\Omega} = (\cos \theta_\ell, \cos \theta_K, \phi)$~\cite{Altmannshofer:2008dz}.
The angular basis used in this work is defined in Ref.~\cite{LHCB-PAPER-2013-019}. 
The five-dimensional differential decay rate ${\rm d}^5 \Gamma / ( {\rm d}q^2 {\rm d}k^2 {\rm d}\vec{\Omega} )$ can be computed from combinations of decay amplitudes~\cite{Bobeth:2017vxj,Descotes-Genon:2019bud,Kruger:2005ep}
\begin{eqnarray}
  \label{eq:amp_dep}
{\mathcal{A}}^{L,R}_{\lambda} &= & {\mathcal{N}}\ \bigg\{  \bigg[ (\mathcal{C}_9 \pm \mathcal{C}^\prime_{9})  
										\mp (\mathcal{C}_{10} \pm   \mathcal{C}^\prime_{10}) \bigg]     \mathcal{F}_{\lambda}(q^2, k^2)      \nonumber   \\   
							&&  \quad\;\;  +  \frac{2 m_b M_B}{q^2}    \bigg[   (\mathcal{C}_7 \pm \mathcal{C}^\prime_{7})  
										 \mathcal{F}_{\lambda}^T(q^2, k^2)   -  16 \pi^2  \frac{M_B}{m_b} \mathcal{H}_{\lambda}(q^2,k^{2}) \bigg]  \bigg\} 
\end{eqnarray}
and corresponding angular terms~\cite{Altmannshofer:2008dz,Kruger:2005ep},
where $\lambda = 0, \parallel, \perp$ refers to the polarisation of the \Kstarz meson,
$L$ and $R$ to the left- and right-hand chirality of the dimuon current, ${\cal{N}}$ is a normalization factor and $m_b$ and $M_{B}$ are the masses of the \textit{b} quark and the \textit{B} meson~\cite{PDG2022}.
Finally, all 
non-perturbative effects are contained within the local, ${\cal{F}}_{\lambda}^{(T)}$, and non-local, ${\cal{H}}_{\lambda}$, form-factors (FFs).
The numerical values for the Wilson coefficients at the $b$-quark mass scale of $\mu_{b}= 4.2 \gevcc$ are calculated in the SM as $\mathcal{C}^{\rm{SM}}_{7} = -0.337, \, \mathcal{C}^{\rm{SM}}_9 = 4.27, \, \mathcal{C}^{\rm{SM}}_{10} = -4.17$, and $\mathcal{C}^{\prime\, \rm{SM}}_{7,9, 10} \simeq 0$~\cite{Bobeth:1999mk,Gorbahn:2004my}.
Local form-factors can be assessed by
light-cone sum rules~\cite{Khodjamirian:2017fxg,Straub:2015ica,Agadjanov:2016fbd,Gubernari:2018wyi,Descotes-Genon:2019bud} 
and lattice QCD~\cite{Gao:2019lta,Bouchard:2013eph,Bailey:2015dka,Horgan:2013hoa,Horgan:2015vla} 
techniques.
Non-local contributions from $b\to c\bar{c}s$ operators
are more difficult to calculate reliably from first principles and only recently a 
rigorous approach that relies on the analytical structure of these matrix elements
has been formulated~\cite{Bobeth:2017vxj,Gubernari:2022hxn}. 
This approach isolates the $\psi_{n}$ resonance poles, where $\psi_{n}$  is a \jpsi or \psitwos state,  and constructs a polynomial expansion for the remaining contributions
in terms of a conformal variable, $z(q^{2})$.
In order to acquire control over the size of the coefficients of the expansion, data on $B^{0}\to \psi_{n}K^{*0}$ decays as well as SM predictions~\cite{Gubernari:2020eft} for the ratios $\mathcal{H}_\lambda / \mathcal{F}_\lambda$  
at negative $q^{2}$ are employed.
In this Letter, the role of the theoretical inputs on the determination of non-local effects (${\cal{H}}_{\lambda}$) is examined, as well as their impact  on the estimation of short-distance physics parameters ($\mathcal{C}_i^{(\prime)}$ and ${\cal{F}}_{\lambda}$).

The observed $\Bz \to K^+ \pi^-  \mu^+\mu^-$ differential decay rate receives a non-negligible contribution from decays where the $K^+\pi^-$ system appears in a scalar (S-wave) configuration~\cite{LHCb-PAPER-2016-012}.
This is taken into account by introducing an additional pair of decay amplitudes and allowing interference and pure S-wave angular terms in the differential decay rate~\cite{Becirevic:2012dp,Lu:2011jm}.
Non-local contributions to the S-wave amplitudes are neglected,
while their normalisation is decoupled from the $\Bz \to \Kstarz  \mu^+\mu^-$ amplitudes by introducing a complex coefficient that describes the relative magnitude and phases to be determined from data.
Finally, the absolute scale of the Wilson coefficients is set by the branching fraction of the decay, which is related to the integral of the differential decay rate over the desired $q^{2}$ and $k^{2}$ ranges through
 \begin{equation}\label{eq:Br_Kstmm}
      \mathcal{B}(B^0 \to K^{*0} \mu^+\mu^-)  = \frac{\tau_B}{\hbar}  \int_{q^2_{\min}}^{q^2_{\max}} \int_{k^2_{\min}}^{k^2_{\max}} \frac{{\textrm{d}}^2 \Gamma}{{\textrm{d}} q^2 {\textrm{d}} k^2} {\textrm{d}} q^2 {\textrm{d}} k^2  \, ,
\end{equation}
where $\tau_B$ is the lifetime of the \Bz meson.

The dataset used in this analysis is the same as employed in Ref.~\cite{LHCB-PAPER-2020-002} and corresponds to an
integrated luminosity of $4.7\,\invfb$
of proton-proton ($pp$) collisions collected with the LHCb experiment during 2011, 2012 and 2016.
The \lhcb detector is a single-arm forward
spectrometer covering the \mbox{pseudorapidity} range $2<\eta <5$, 
described in detail in Refs.~\cite{LHCb-DP-2008-001,LHCb-DP-2014-002}. 
Samples of simulated events produced with the software described in Refs.~\cite{Sjostrand:2007gs,Lange:2001uf,Allison:2006ve, Agostinelli:2002hh}
are used to determine the reconstruction and selection efficiencies for signal
candidates and to estimate the contamination from residual backgrounds. 
Inaccuracies in the simulations are corrected for using control samples from data.

The $\Kstarz$ candidates are selected with $K^{+}\pi^-$ invariant mass within $100\mevcc$ of the known $\Kstarz$ mass~\cite{PDG2022}.
Candidates with $q^2 < 1.1$ or $q^2 > 15.0\gevgevcccc$ are excluded, to remove contributions from light-quark resonances and from charmonium states beyond the open-charm threshold, respectively.
Signal candidates are selected in two $q^2$ regions, $[1.1, \, 8.0]$ and $[11.0,\, 12.5]\gevgevcccc$, while
the tree-level decays $\Bz \to \jpsi \Kstarz$ and \mbox{$\Bz \to \psitwos \Kstarz$}, where the charmonium resonance decays to \mumu, 
are retained as control regions in the intervals $[8.0, \, 11.0]$ and $[12.5,\, 15.0]\gevgevcccc$, respectively, in order to validate several procedures of the analysis.
A total of $2568 \pm 60$ signal decays and approximately $ 677\,000$  $\Bz \to \jpsi \Kstarz$  and $43\,700$ $\Bz \to \psitwos \Kstarz$ decays are selected. 

The variation of the efficiency to reconstruct and select the signal  across the kinematic phase space is accounted for in the fit by multiplying the differential decay rate by an efficiency function obtained from simulated samples.
This efficiency function is parameterised 
using different orders of Legendre polynomials, each depending on one angle or $q^2$, without assuming factorisation.
No significant dependence of the efficiency on $k^2$ is observed.
Moreover, the relative efficiency between rare and control modes is obtained from these simulations and
the efficiency model is validated by comparing the measured ratio of branching fractions of the two control modes  to its known value~\cite{PDG2022}.

An extended unbinned maximum-likelihood fit to the five-dimensional differential decay rate, in $q^{2}, \, k^2$ and the three decay angles, and the \textit{B}-candidate invariant mass distribution, $m_{K\pi\mu\mu}$, is performed using the \textsc{TensorFlow} library~\cite{tensorflow2015-whitepaper} with an interface to the \textsc{Minuit} minimisation algorithm~\cite{James:1975dr,Brun:1997pa}. 
The $m_{K\pi\mu\mu}$ distribution is used in the fit to discriminate signal from background, where the background is composed primarily from random combination of tracks.
The real parts of the ${\mathcal{C}}^{(\prime)}_9$ and ${\mathcal{C}}^{(\prime)}_{10}$ coefficients are allowed to vary in the fit, while the coefficients ${\mathcal{C}}^{(\prime)}_{7}$, which are strongly constrained by radiative $B$ decays~\cite{Paul:2016urs}, are fixed to their SM values.
Since \Bz and \Bzb decays are treated jointly in the analysis, only the \textit{CP}-averaged decay rate is accessed and no sensitivity to the
imaginary part of the Wilson coefficients can be achieved.
In addition to the above-mentioned Wilson coefficients, a large number of signal parameters is extracted simultaneously from the fit to data: the local and non-local FF parameters, the Cabibbo-Kobayashi-Maskawa (CKM) factor $V_{tb}V^*_{ts}$ that enters in the normalisation of Eq.~\ref{eq:amp_dep} and the S-wave relative magnitude and phase, while the local S-wave FFs are treated as nuisance parameters.

The convergence of the polynomial expansion employed to model the non-local hadronic contributions $\mathcal{H}_\lambda$ is carefully investigated.
The polynomial expansion is performed around $q^{2} = 0$,
the truncation point of the expansion is chosen by repeating the fit with increasing orders of polynomials 
and the Akaike information criterion~\cite{Akaike} is used to decide on the statistical relevance of each additional set of coefficients. 
A fourth-order expansion is found to be sufficient. 
The \textit{B}-candidate mass distribution for the signal is parameterised by a sum of two 
Crystal Ball functions~\cite{Skwarnicki:1986xj}.
Finally, the $k^{2}$ dependence of the signal amplitudes is modelled using a relativistic Breit--Wigner function for the spin-1 \Kstarz resonance and the LASS parameterisation~\cite{Aston:1987ir} for
the S-wave component.

The background is modelled
by second-order polynomial functions of the decay angles and $q^2$, whose coefficients are allowed to vary in the fit.
The $k^2$ distribution is described 
by the sum of a linear function and a Breit--Wigner amplitude squared, where the former accounts for a pure combinatorial component and the latter accommodates 
genuine $K^{*0}$ resonances associated with random \mup and \mun tracks. 
The reconstructed $B$ mass distribution of the background is parameterised by an exponential function. 
A significant sculpting of $\cos \thetak$ as a function of $q^2$ and $m_{K\pi\mu\mu}$ is observed due to a kinematic veto used to reject \mbox{$\Bp \to K^+ \mumu$} decays.
This distortion is accounted for by multiplying the combinatorial background parameterisation by a three-dimensional efficiency correction factor obtained from a $B^0 \to K^{*0} e^\pm \mu^\mp$ control sample.

The observed signal yield determined by the extended fit is expressed in terms of the branching fraction of the decay given in Eq.~\ref{eq:Br_Kstmm} through
\begin{equation}\label{eq:nsig}
  N_{\rm sig} = N_{\jpsi K \pi} \times \frac{\mathcal{B}(B^0 \to \Kstarz \mumu) \times \frac{2}{3}  }{\mathcal{B}(B^0 \to \jpsi K^+ \pi^-)  \times f^{\jpsi K \pi} \times \mathcal{B}(\jpsi \to \mu^+\mu^-)  } 
  \times R_\varepsilon \,,
\end{equation}
where 
the factor $\frac{2}{3}$ comes from the $\Kstarz \to K^+ \pi^-$ decay probability and
the signal branching fraction is normalised to the $B^0 \to \jpsi K^+ \pi^-$ control channel in order to reduce the associated systematic uncertainty.
The yield of the control channel, $N_{\jpsi K \pi}$, is obtained directly from a mass fit, while
the resonant and charmonium branching ratios $\mathcal{B}(\Bz \to \jpsi K^+ \pi^-) = (1.15 \pm 0.01 \pm 0.05) \times 10^{-3}$ and $\mathcal{B}(\jpsi \to \mumu) = (5.96 \pm 0.03) \times 10^{-2}$ are taken from Refs.~\cite{Chilikin:2014bkk} and~\cite{PDG2022}, respectively.
The numerical factor $f^{\jpsi K \pi}= 0.644 \pm 0.010$
scales the total $B\to \jpsi K^+ \pi^-$ branching ratio in the $k^{2}$ range of this analysis~\cite{LHCb-PAPER-2023-032}
and $R_{\varepsilon}$ is the relative efficiency between the signal and control modes obtained from simulated samples.

A set of external constraints is  imposed on the signal model to ensure the stability of the amplitude fit in a similar fashion to Refs.~\cite{Gubernari:2022hxn,Chrzaszcz:2018yza}: 
the CKM elements $V_{tb}$ and $V^*_{ts}$
are constrained to the values obtained from the SM fit of the Unitarity triangle~\cite{Charles:2004jd};
the local FFs  for the $\Bz \to \Kstarz$ transition are constrained to a combination~\cite{Gubernari:2022hxn,Gubernari:2023puw} of
light-cone sum rules~\cite{Gubernari:2018wyi,Descotes-Genon:2019bud} 
and lattice QCD~\cite{Horgan:2015vla} 
results; while for the S-wave amplitudes the local FFs are assumed to have the same $q^{2}$ dependence as in $B^+ \to K^+$ transitions and are
constrained to the results of Ref.~\cite{Bailey:2015dka} but have their uncertainties inflated by a factor of three to account for the different meson species.
An alternative choice of fixing the S-wave local FFs to the values from Ref.~\cite{Doring:2013wka} is considered as a source of systematic uncertainty.
The magnitudes and phase differences of the 
resonant amplitudes for the $\Bz \to \psi_{n}\Kstarz$ decays are instrumental to constrain the values of $\mathcal{H}_\lambda$ at the $\jpsi$ and $\psitwos$ resonance poles.
These are taken from measurements by both LHCb and $B$-factory experiments~\cite{Aubert:2007hz,Chilikin:2013tch,Chilikin:2014bkk,LHCb-PAPER-2013-023,LHCb-PAPER-2012-010}. 
Finally, the SM predictions for the real and imaginary parts of the ratio  $\mathcal{H}_\lambda / \mathcal{F}_\lambda$ in the negative-$q^{2}$ region are taken from Refs.~\cite{Gubernari:2020eft,Gubernari:2022hxn} and are used as constraints in the fit.

Systematic uncertainties are studied using pseudoexperiments in which one or more  parameters are varied and the values of interest are determined with and without this variation. 
The difference between the two sets of results  is then taken as the associated
systematic uncertainty.
The total systematic uncertainty varies between 15 and $35\,\%$ of the statistical uncertainty,
depending on the considered Wilson coefficient.
The dominant sources of systematic uncertainties are associated with the external branching fraction measurements entering in Eq.~\ref{eq:nsig}, where all the external inputs are varied within their uncertainties. 
The uncertainty on $R_\varepsilon$ originates from the finite size of the simulated samples and from the model dependence of the simulation. 
The uncertainty on $f^{\jpsi K \pi}$ is determined from pseudoexperiments generated from the amplitude model of Ref.~\cite{Chilikin:2014bkk}. 
The systematic uncertainty due to ignoring non-local effects in the scalar amplitudes is assessed by generating pseudoexperiments where non-local FFs are assumed to be of the same order of those observed in the longitudinal P-wave amplitude. 
All other considered sources of systematic uncertainty are negligible.

\begin{figure}[!t]
  \begin{center}
    \includegraphics[width=.49\textwidth]{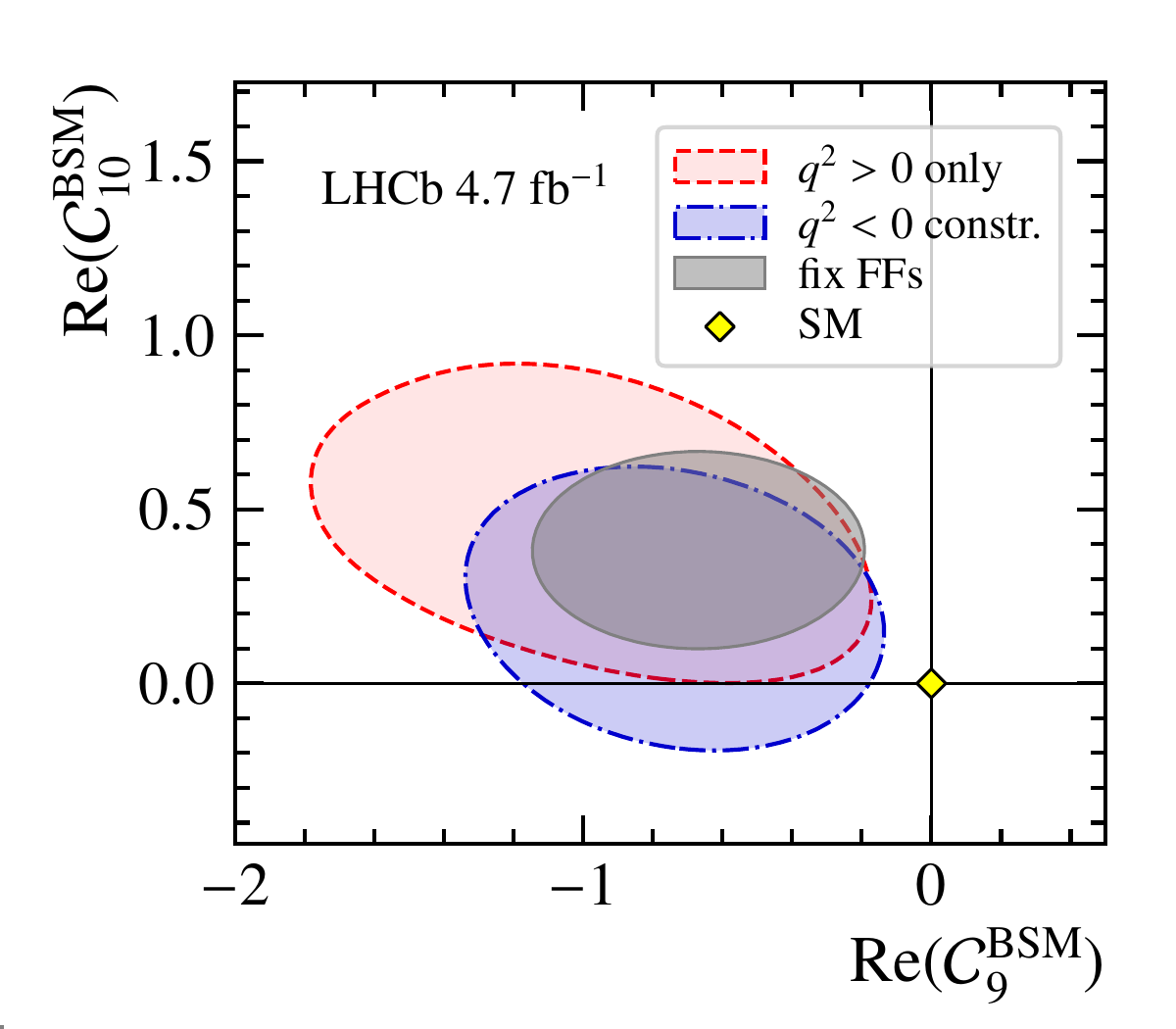}
    \includegraphics[width=.49\textwidth]{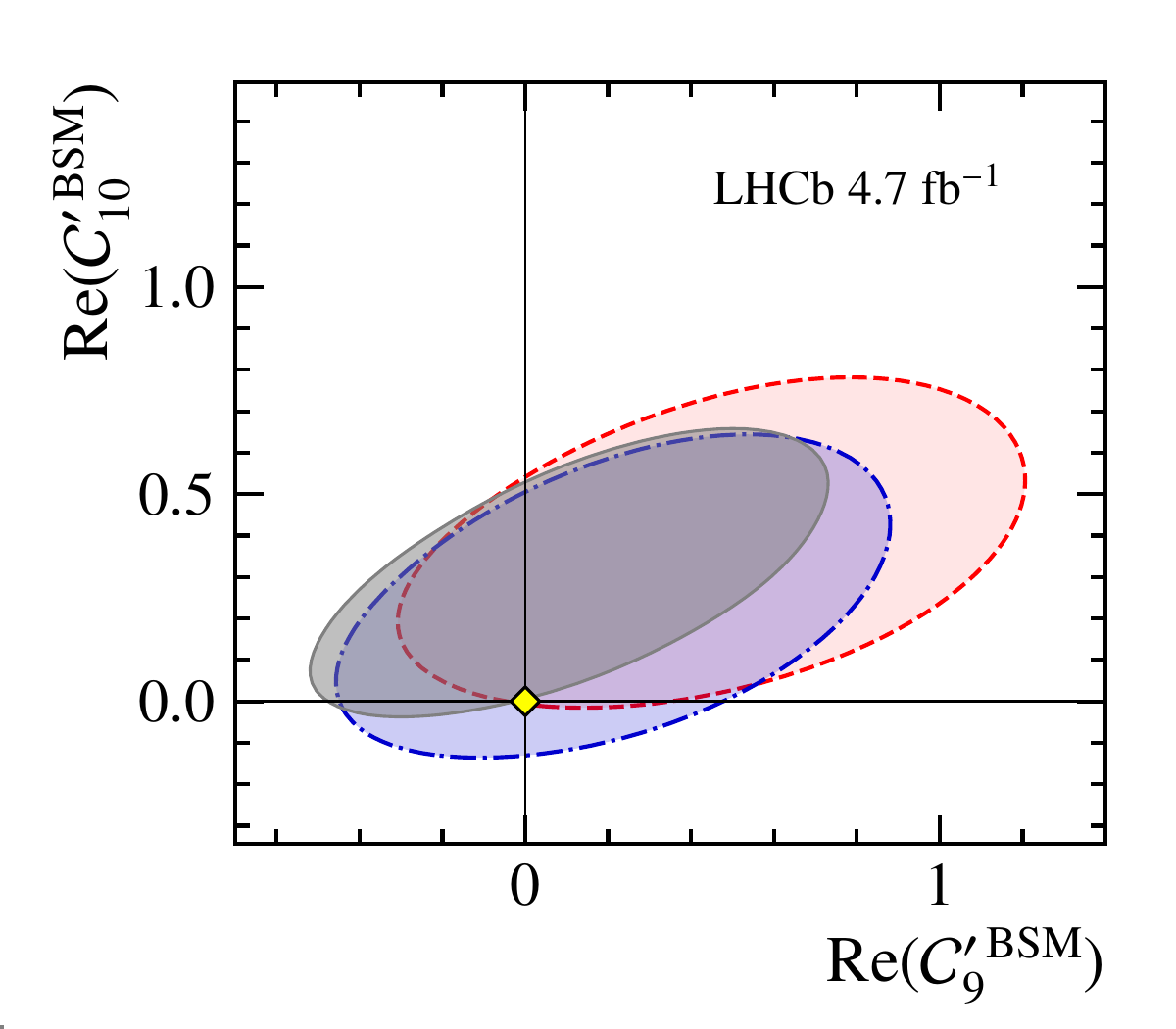}
  \end{center}
  \vspace{-6mm}
  \caption{Two-dimensional profile-likelihood contours of the (left) $\mathcal{C}_9^{\mathrm{BSM}}$--$\mathcal{C}_{10}^{\mathrm{BSM}}$ and (right) $\mathcal{C}_9^{\prime \, \mathrm{BSM}}$--$\mathcal{C}_{10}^{\prime \, \mathrm{BSM}}$ pairs of Wilson coefficients at $68\%$ confidence level 
  with (blue) and without (red) SM constraints at $q^{2}<0$. The fit is also repeated considering  local FFs to be fixed to their SM  predictions (grey)~\cite{Gubernari:2018wyi,Gubernari:2022hxn,Horgan:2015vla}.}
  \label{fig:WCsFits}
\end{figure}

The results of the fit to the data are shown in Fig.~\ref{fig:WCsFits} (blue contours) as two-dimensional profile-likelihood projections for $\mathcal{C}_9^{(\prime)\, \mathrm{BSM}}$ and $\mathcal{C}_{10}^{(\prime)\, \mathrm{BSM}}$, where the superscript BSM indicates a difference with respect to the SM predictions.
When allowing for non-local hadronic effects,
the fit results still yield a $\mathcal{C}_9$ value that is somewhat different from the SM prediction, 
however, the global significance of the differences in the Wilson coefficients is equivalent to 1.4 standard deviations ($\sigma$), considering both statistical and systematic uncertainties.
In order to evaluate the compatibility of each Wilson coefficient with the SM, one-dimensional profile likelihood scans  are performed on the individual coefficients.
The largest deviation is associated to a shift in $\mathcal{C}_9$ of $-0.7$ with a significance of $1.8\,\sigma$.
These results show a good qualitative agreement with global analyses of $b \to s \mu^+ \mu^-$ transitions~\cite{Gubernari:2022hxn,Greljo:2022jac,Alguero:2023jeh,Ciuchini:2022wbq}.
In comparison with the existing literature, the present analysis relies on the unbinned use of the experimental data, a $z$-expansion for the treatment of non-local contribution (as in Ref.~\cite{Gubernari:2022hxn}) and  focus on only $\Bz\to\Kstarz\mumu$ data, while global analyses typically include data on other $b \to s\mumu$ processes.

The impact of long-distance contributions on the determination of the genuine short-distance effects can be better investigated by repeating the amplitude fit using alternative theory assumptions, as overlaid in Fig.~\ref{fig:WCsFits}. 
As a first test, the theory constraints at negative $q^{2}$ are removed from the fit. 
In this case, a second-order polynomial is sufficient to accommodate the non-local FF contributions and a point within the charmonium resonance region is used as a reference for the expansion.
A similar compatibility to the SM is observed but with a larger statistical uncertainty (red contours). 
Another interesting behaviour is observed in the role of the local FFs in the determination of the non-local effects. 
Since  the largest uncertainty on the theory prediction of $\mathcal{H}_\lambda$ at negative $q^2$ is due to the local FF uncertainties~\cite{Gubernari:2020eft,Gubernari:2022hxn},
there is intrinsically a strong correlation in the fit between the local and non-local parameters. 
As a result, 
removing the theory constraints at $q^2 <0$  has an effect on the determination of the local FFs from the fit and, in turn, on all the Wilson coefficients.
An overall shift in all the coefficients is  observed between the two fit results.
This behaviour is further studied by repeating the default fit with fixed local FF values (gray contours in Fig.~\ref{fig:WCsFits}).
This artificial configuration, which assumes perfect knowledge of $\mathcal{F}_\lambda^{(T)}$, 
illustrates how the uncertainties associated to the local FFs dominate and prevent a clean extraction of $\mathcal{C}_9$ and $\mathcal{C}_{10}$.

\begin{figure}[!t]
\begin{center}
    \includegraphics[width=0.62\textwidth]{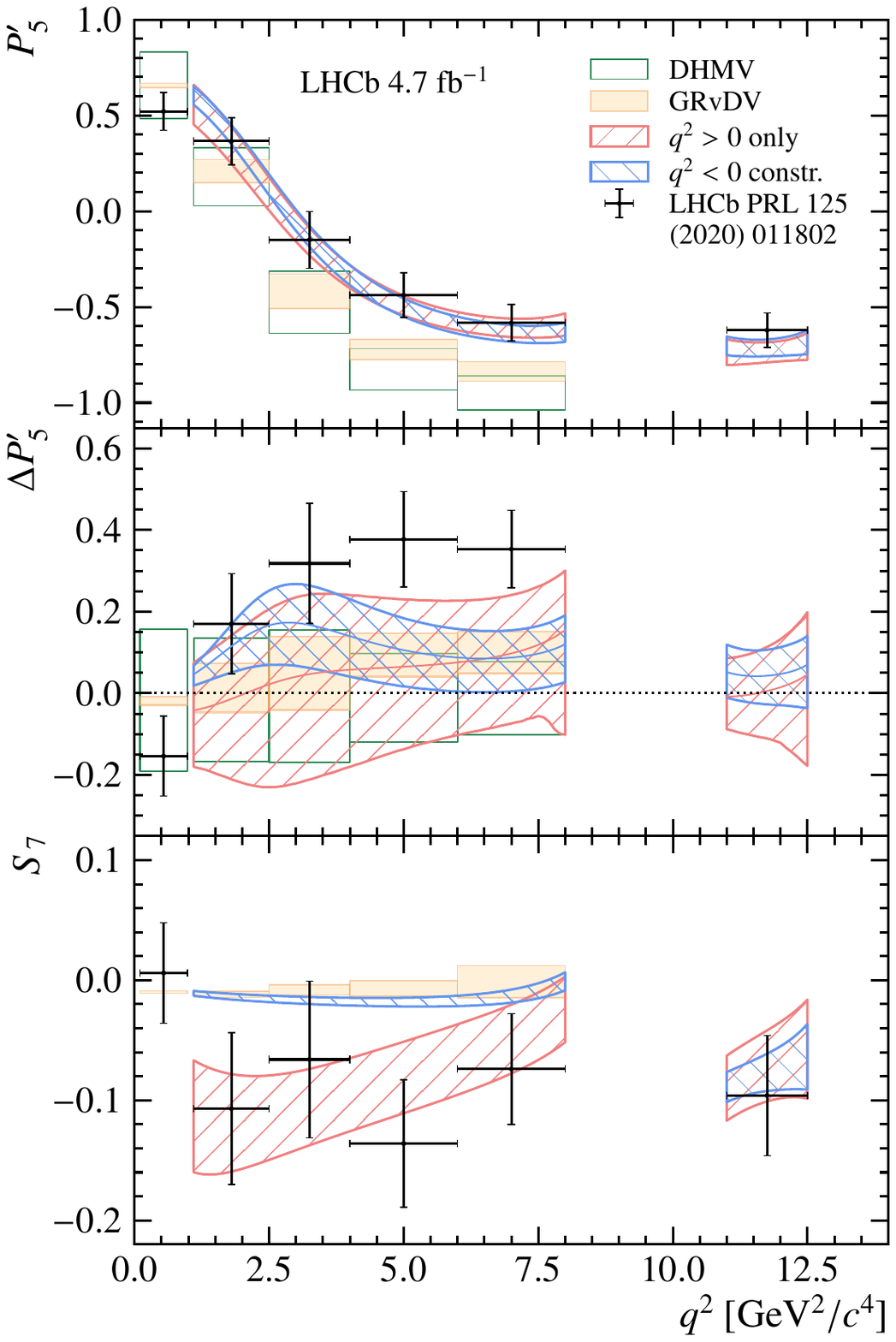}
\end{center}
  \vspace{-6mm}
  \caption{
    Determination of the (top) $P^{\prime}_{5}$ and (bottom) $S_7$ angular observables
    obtained from the amplitude fit. 
    Results are shown for fits  with (blue) and  without (red) constraints from $q^{2} < 0$, at 68\% confidence level.
  The LHCb binned angular analysis~\cite{LHCB-PAPER-2020-002} (black dots) and SM predictions from DHMV~\cite{Descotes-Genon:2014uoa,Khodjamirian:2010vf} (green) and GRvDV~\cite{Gubernari:2018wyi,Gubernari:2022hxn,Horgan:2015vla} (yellow) are overlaid for reference. 
  The center panel shows the non-local contributions to $P^{\prime}_{5}$ determined from the amplitude fit.
  Here the difference between the LHCb binned angular analysis~\cite{LHCB-PAPER-2020-002} and SM central value from DHMV is overlaid for reference (black dots), together with the uncertainty on the DHMV SM prediction (green) and difference between the SM predictions from GRvDV and DHMV (yellow). 
    Theoretical predictions are shown only for $q^2 < 8\gevgevcccc$.
  }
  \label{fig:P5p:and:S7}
\end{figure}

Figure~\ref{fig:P5p:and:S7} (top) reports the determination of the angular observable $P_5^\prime$ obtained from the full amplitude fit compared to the result of the previous binned angular analysis~\cite{LHCB-PAPER-2020-002} and different sets of SM predictions~\cite{Descotes-Genon:2014uoa,Khodjamirian:2010vf,Gubernari:2018wyi,Gubernari:2022hxn,Horgan:2015vla}.
A good consistency is found between the results of the binned and unbinned analyses.
Figure~\ref{fig:P5p:and:S7} (centre)  illustrates the contribution due to non-local
hadronic effects to the $P^{\prime}_{5}$ observable obtained in this analysis. 
This contribution is isolated by varying the non-local parameters within their uncertainties and subtracting the resulting value of the observable by the same quantity re-evaluated with the non-local parameters fixed to zero. 
Non-local effects are found to contribute with a small positive shift to the $P_5^\prime$ observable in a direction that reduces the discrepancy between the SM and the data. 
The constraints imposed by the $q^2<0$ predictions significantly reduce the uncertainty on the non-local effects.
In general, the impact of non-local contributions to the different angular observables is found to be consistent whether $q^{2}<0$ constraints are included or not in the fit.
The only exception is the $S_{7}$  observable in Fig.~\ref{fig:P5p:and:S7} (bottom).
Since a non-zero value of $S_{7}$ can only result from a strong phase difference between amplitudes, the observed trend indicates potentially large phase differences in data that can not be accounted for if the theory constraints at $q^2 <0$ are applied.
For completeness, the impact of the non-local effects on all the other angular observables is given in the Supplemental Material~\cite{Suppl}.

In summary, using $pp$ collision data collected with the LHCb experiment
between 2011 and 2016, 
a direct experimental determination of the short-distance contributions in the
$\Bz\to \Kstarz\mu^+\mu^-$ decay is obtained for the first time, 
together with the most accurate characterisation to date of the impact of long-distance effects on the decay process.
The results are consistent with the pattern of modifications to the Wilson coefficients suggested by global analyses of $b \to s \mumu$ processes, but 
the explicit inclusion of non-local contributions in the signal amplitude model is found to reduce the previously observed tension in the $\Bz\to \Kstarz\mu^+\mu^-$ decay to a level below two standard deviations.

\section*{Acknowledgements}
%
%
\noindent 
We are very grateful to Nico Gubernari, M\'{e}ril Reboud, Danny van Dyk, and Javier Virto for the many helpful discussions.
We express our gratitude to our colleagues in the CERN
accelerator departments for the excellent performance of the LHC. We
thank the technical and administrative staff at the LHCb
institutes.
We acknowledge support from CERN and from the national agencies:
CAPES, CNPq, FAPERJ and FINEP (Brazil); 
MOST and NSFC (China); 
CNRS/IN2P3 (France); 
BMBF, DFG and MPG (Germany); 
INFN (Italy); 
NWO (Netherlands); 
MNiSW and NCN (Poland); 
MCID/IFA (Romania); 
MICINN (Spain); 
SNSF and SER (Switzerland); 
NASU (Ukraine); 
STFC (United Kingdom); 
DOE NP and NSF (USA).
We acknowledge the computing resources that are provided by CERN, IN2P3
(France), KIT and DESY (Germany), INFN (Italy), SURF (Netherlands),
PIC (Spain), GridPP (United Kingdom), 
CSCS (Switzerland), IFIN-HH (Romania), CBPF (Brazil),
and Polish WLCG (Poland).
We are indebted to the communities behind the multiple open-source
software packages on which we depend.
Individual groups or members have received support from
ARC and ARDC (Australia);
Key Research Program of Frontier Sciences of CAS, CAS PIFI, CAS CCEPP, 
Fundamental Research Funds for the Central Universities, 
and Sci. \& Tech. Program of Guangzhou (China);
Minciencias (Colombia);
EPLANET, Marie Sk\l{}odowska-Curie Actions, ERC and NextGenerationEU (European Union);
A*MIDEX, ANR, IPhU and Labex P2IO, and R\'{e}gion Auvergne-Rh\^{o}ne-Alpes (France);
AvH Foundation (Germany);
ICSC (Italy); 
GVA, XuntaGal, GENCAT, Inditex, InTalent and Prog.~Atracci\'on Talento, CM (Spain);
SRC (Sweden);
the Leverhulme Trust, the Royal Society
 and UKRI (United Kingdom).

\section*{Supplemental material}
\label{sec:Supplementary}

\begin{figure}[ht]
\begin{center}
    \includegraphics[width=0.45\textwidth]{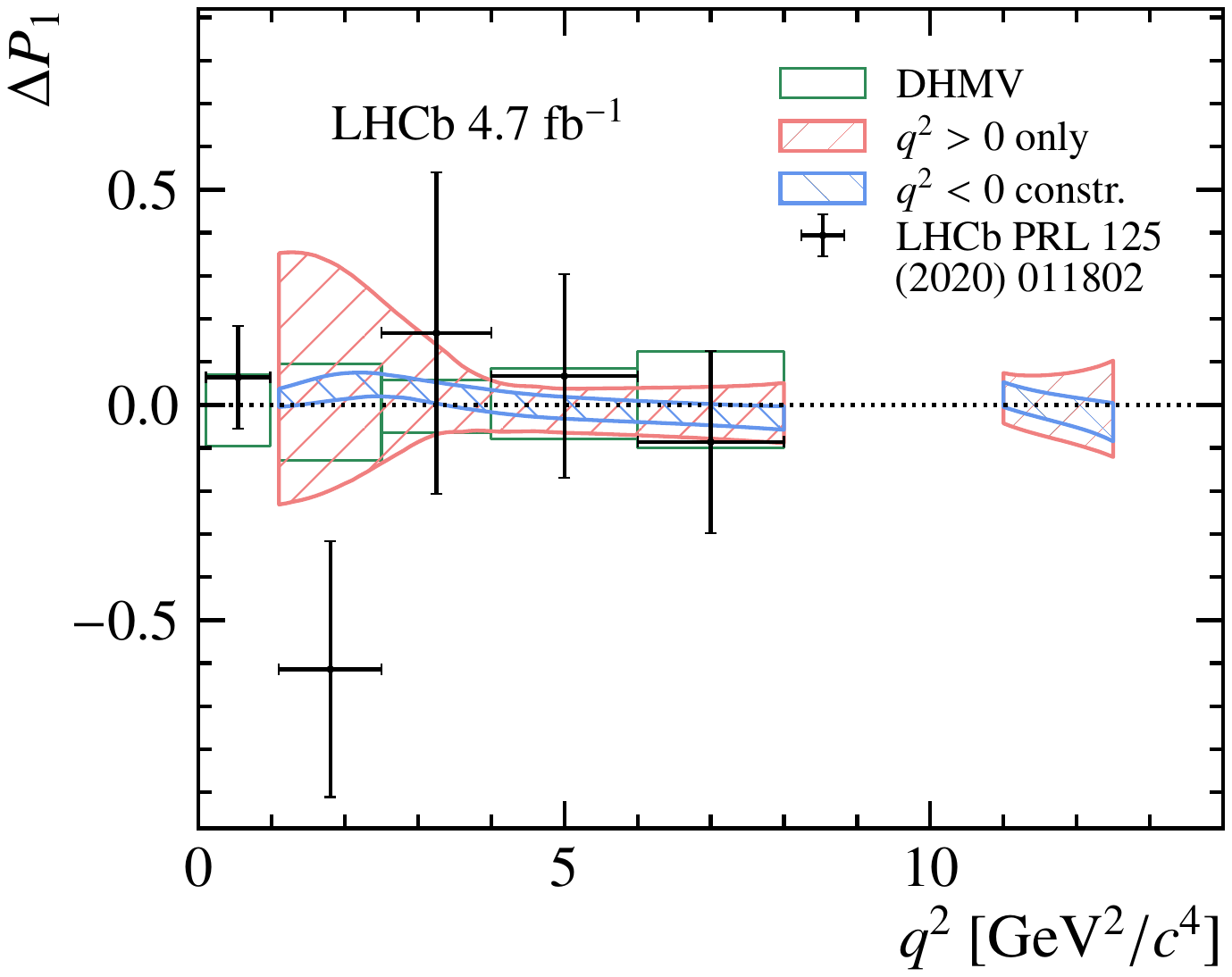}
    \hspace{-10mm}
    \includegraphics[width=0.45\textwidth]{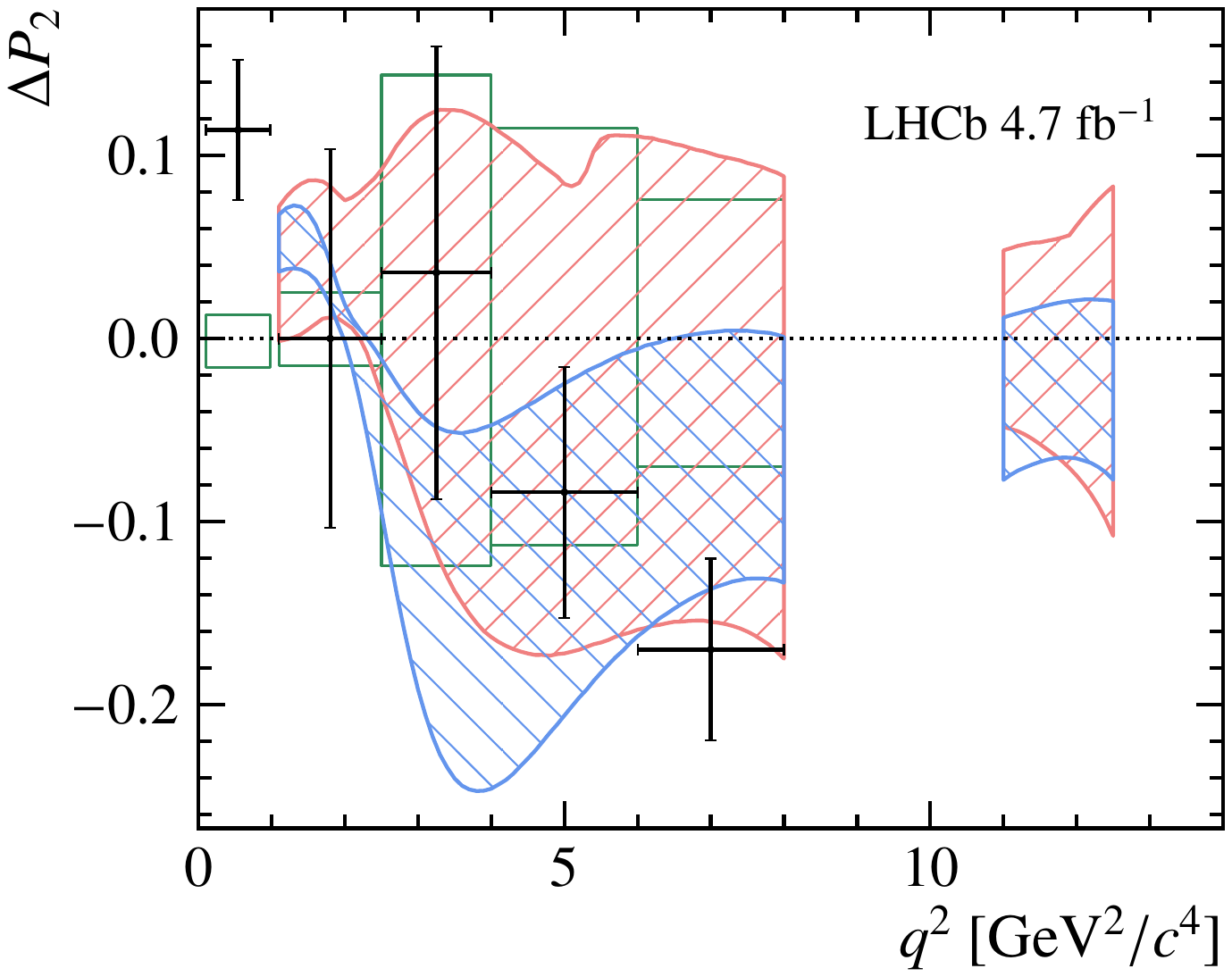}\\
    \vspace{-2.5mm}
    \includegraphics[width=0.45\textwidth]{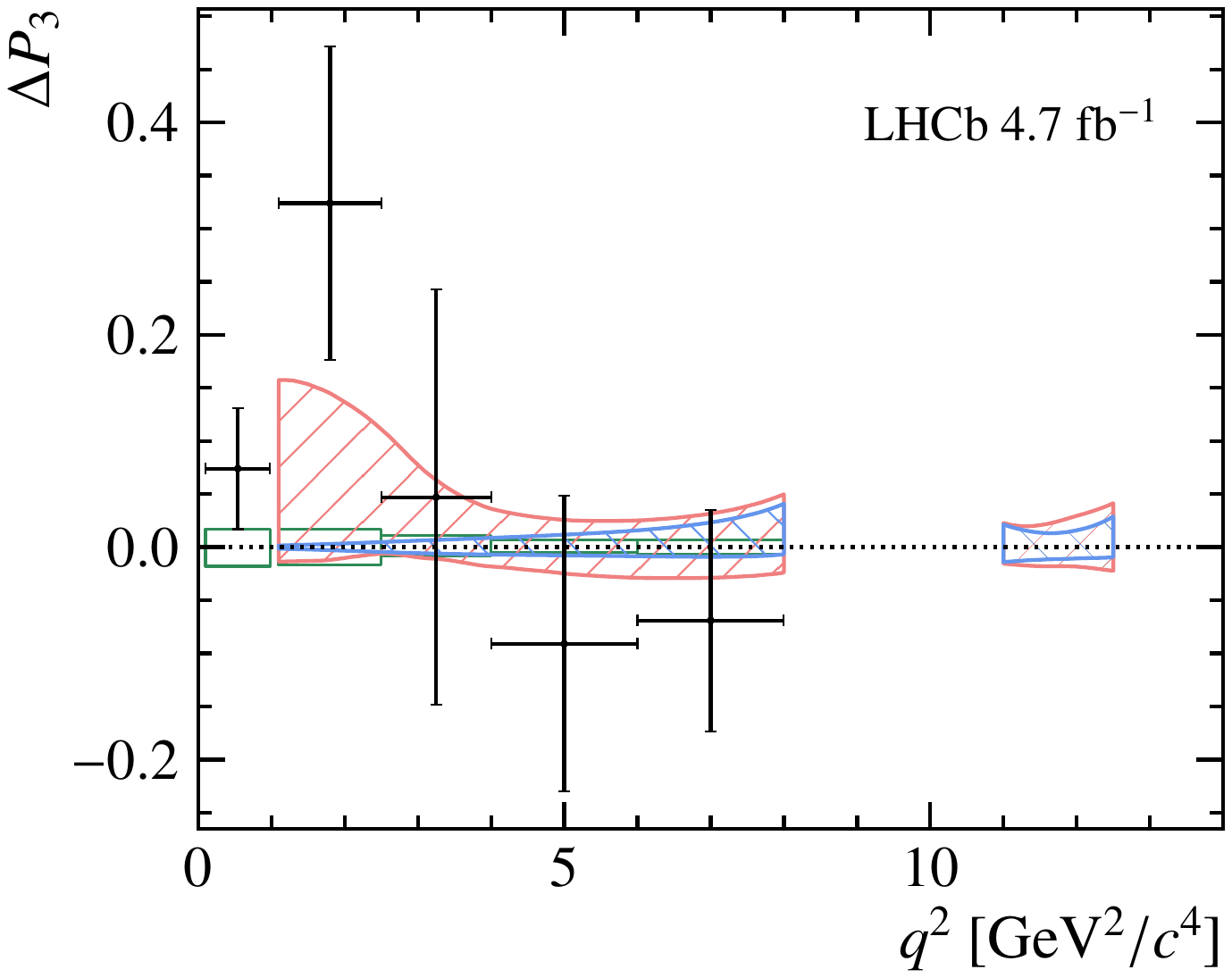}
    \hspace{-10mm}
    \includegraphics[width=0.45\textwidth]{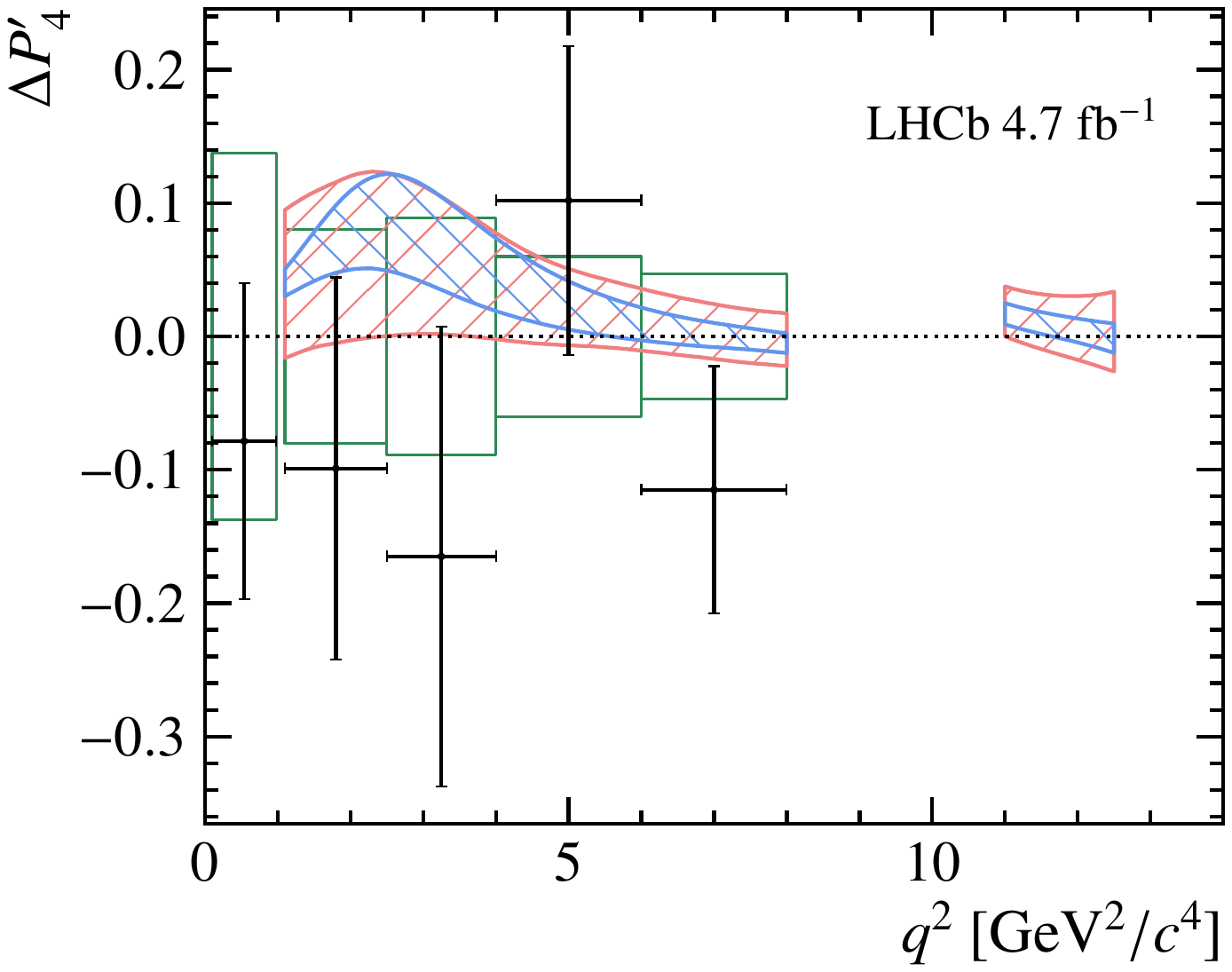}\\
    \vspace{-2.5mm}
    \includegraphics[width=0.45\textwidth]{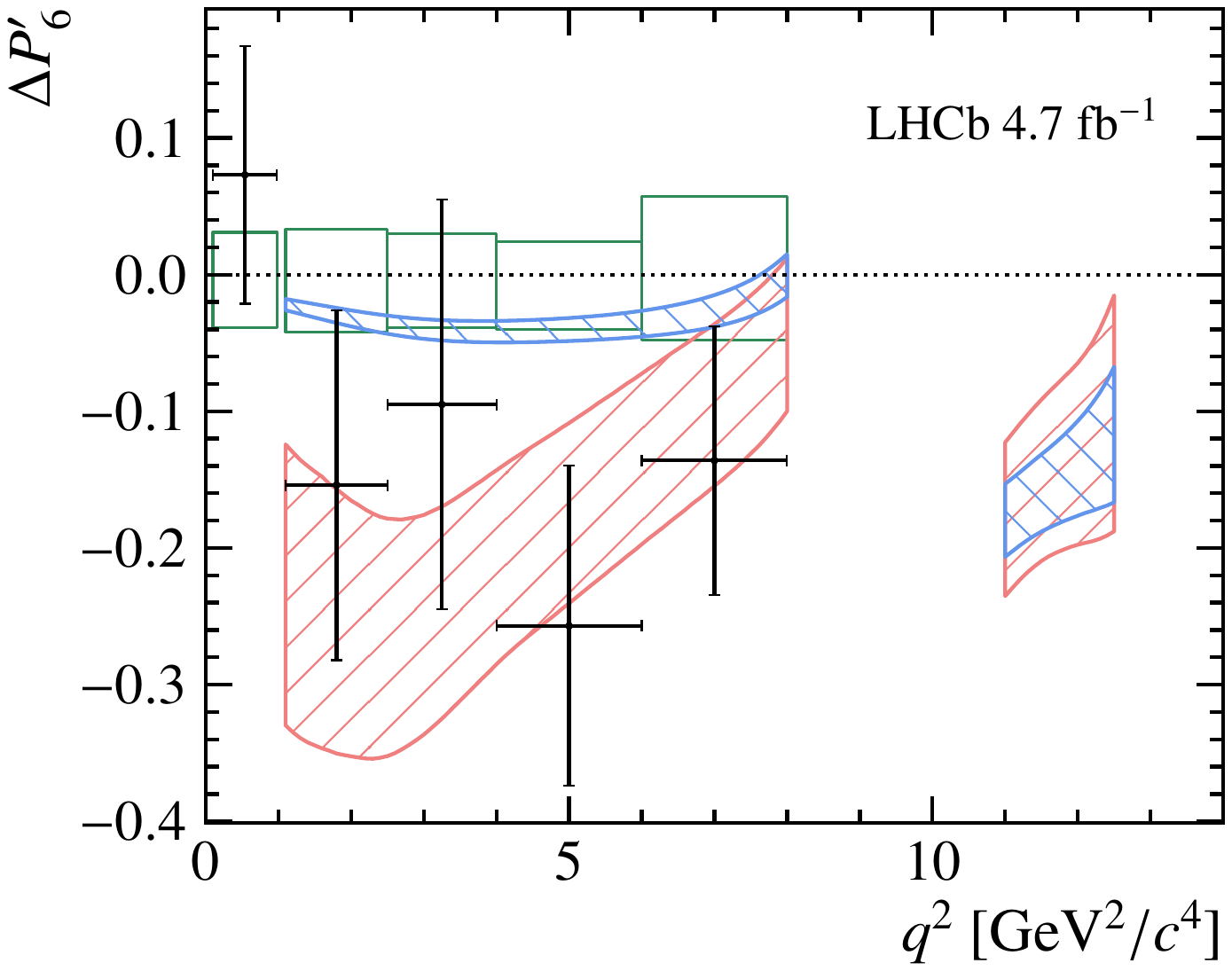}
    \hspace{-10mm}
    \includegraphics[width=0.45\textwidth]{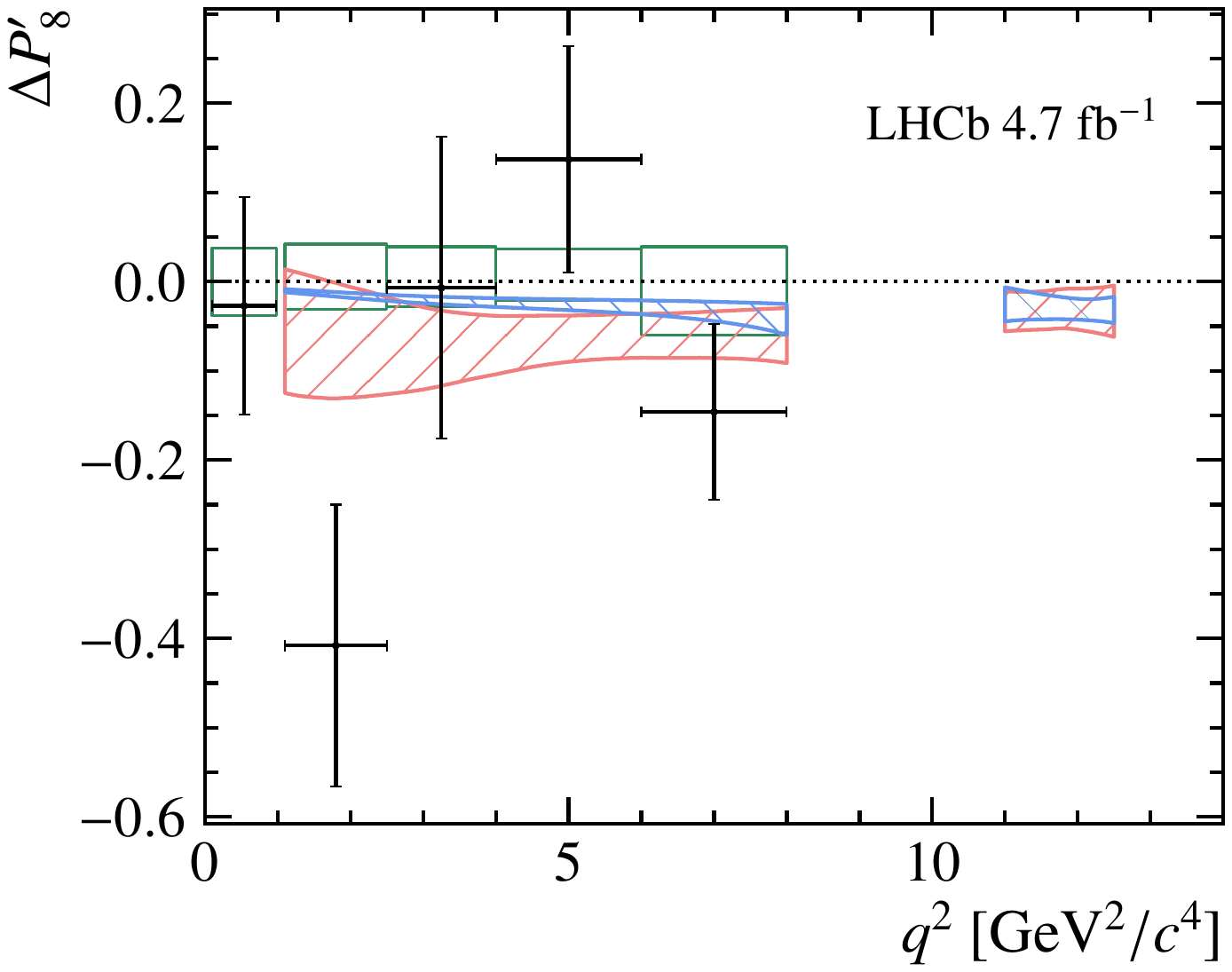}
\end{center}
  \vspace{-6mm}
  \caption{
  Non-local contributions to the angular observables (P-basis) obtained from data fits with (blue) and without (red) constraints from $q^{2} < 0$. The shaded regions are shown at 68\% confidence level. 
  When available, the difference between the LHCb binned angular analysis~\cite{LHCB-PAPER-2020-002}  
  and SM central-value from Refs.~\cite{Descotes-Genon:2014uoa,Khodjamirian:2010vf} is overlaid for reference (black dots), together with the uncertainty on the SM prediction (green).
  }
  \label{fig:Pbasis}
\end{figure}

\begin{figure}[ht]
\begin{center}
    \includegraphics[width=0.45\textwidth]{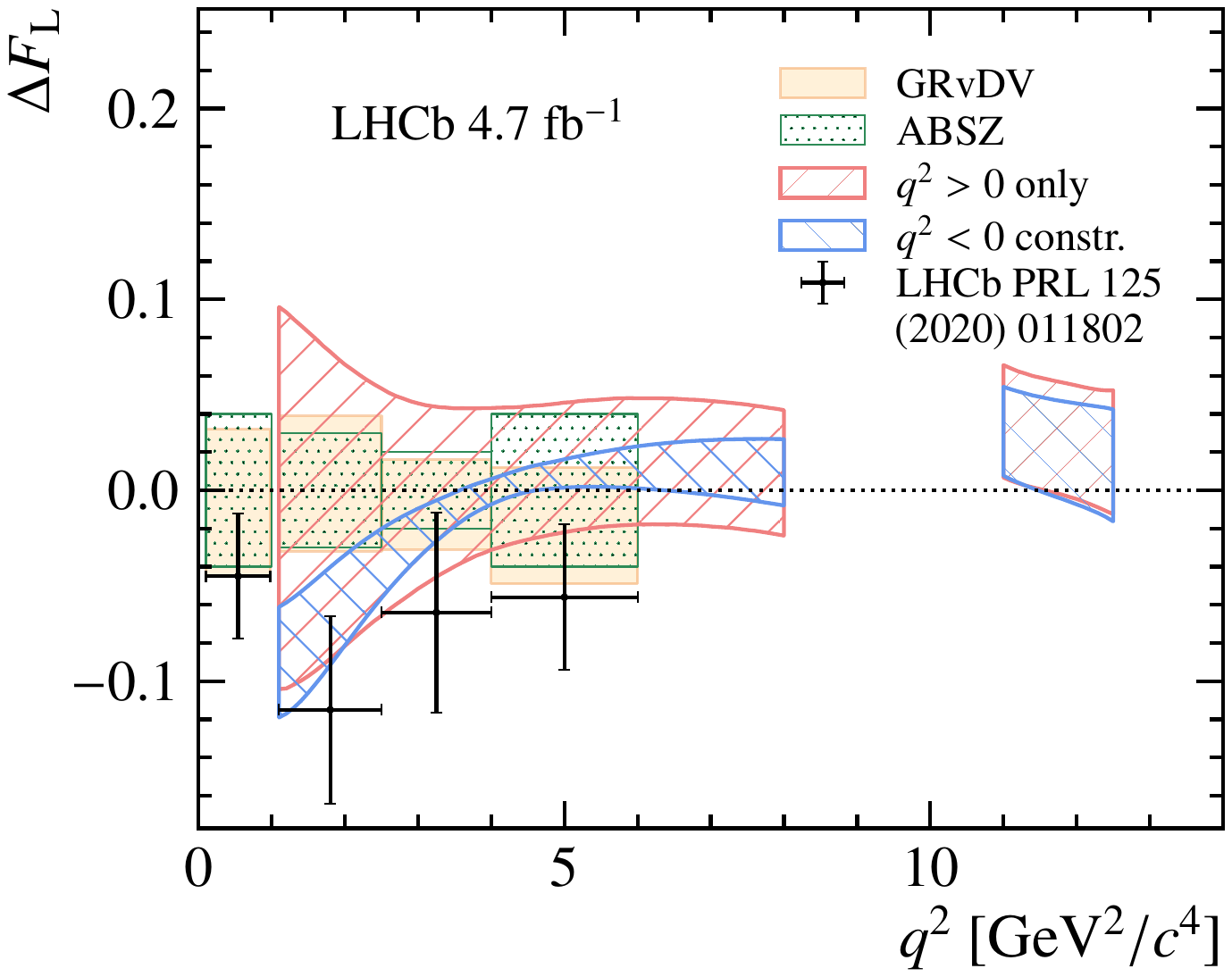}
    \hspace{-10mm}
    \includegraphics[width=0.45\textwidth]{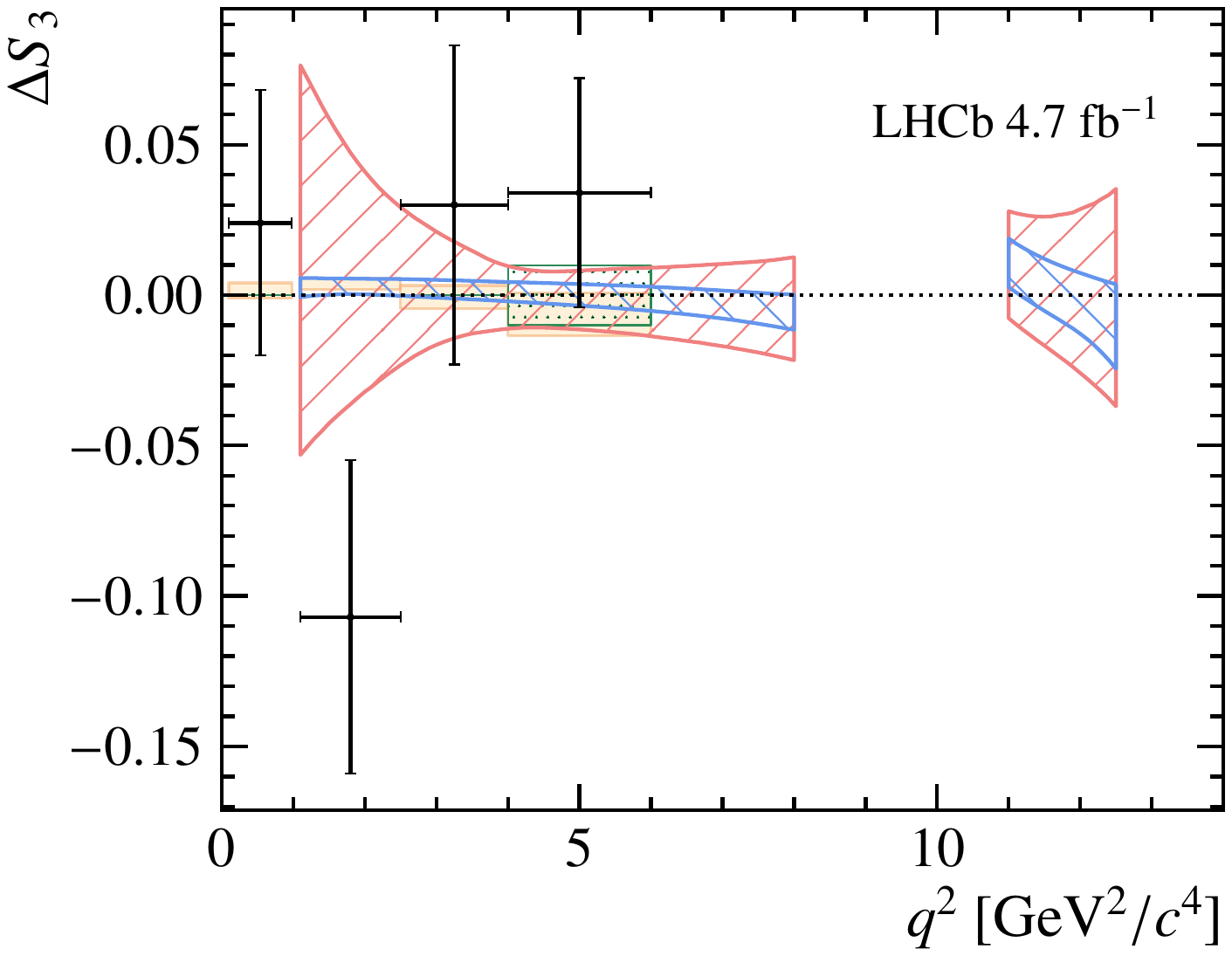}\\
    \vspace{-2.5mm}
    \includegraphics[width=0.45\textwidth]{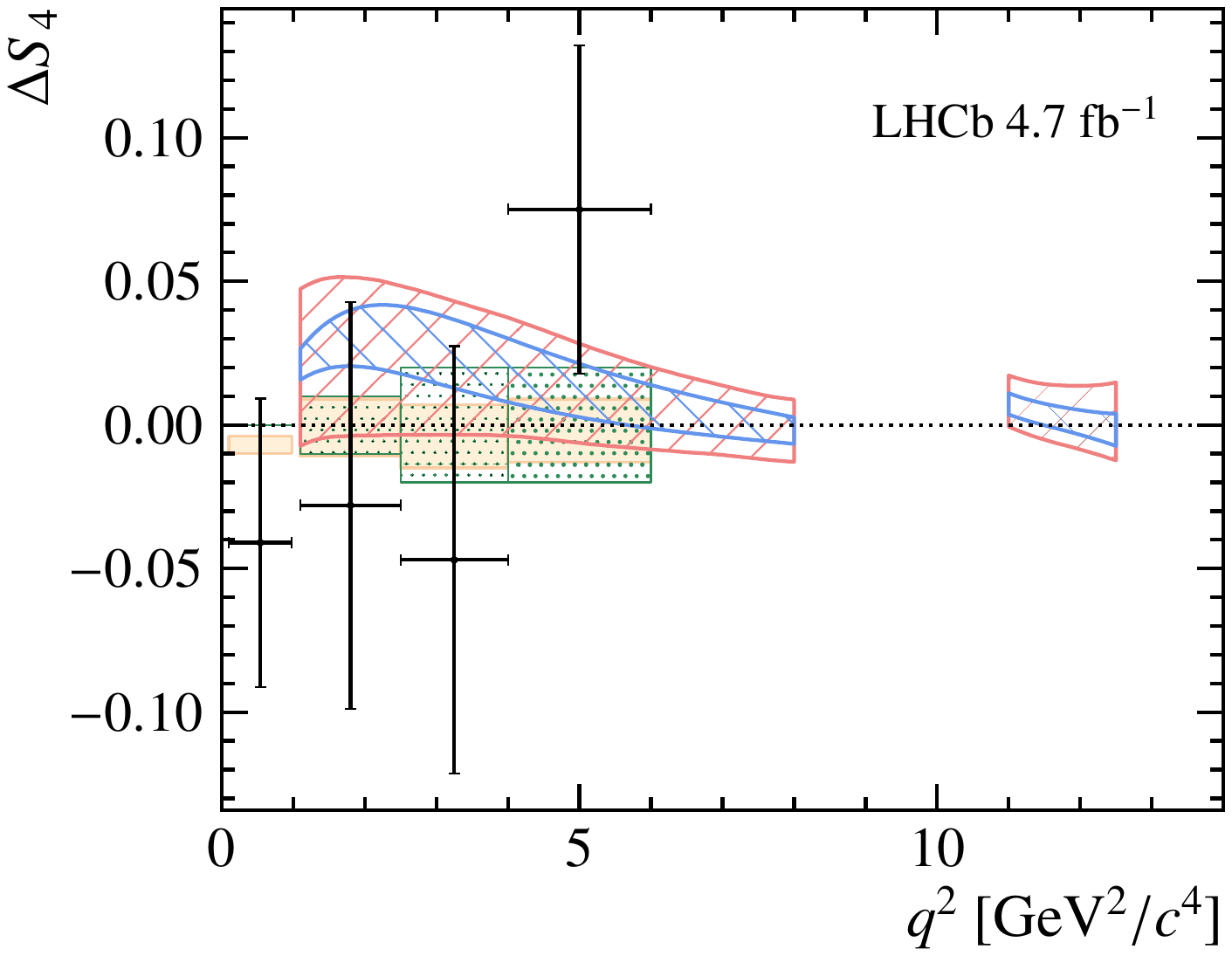}
    \hspace{-10mm}
    \includegraphics[width=0.45\textwidth]{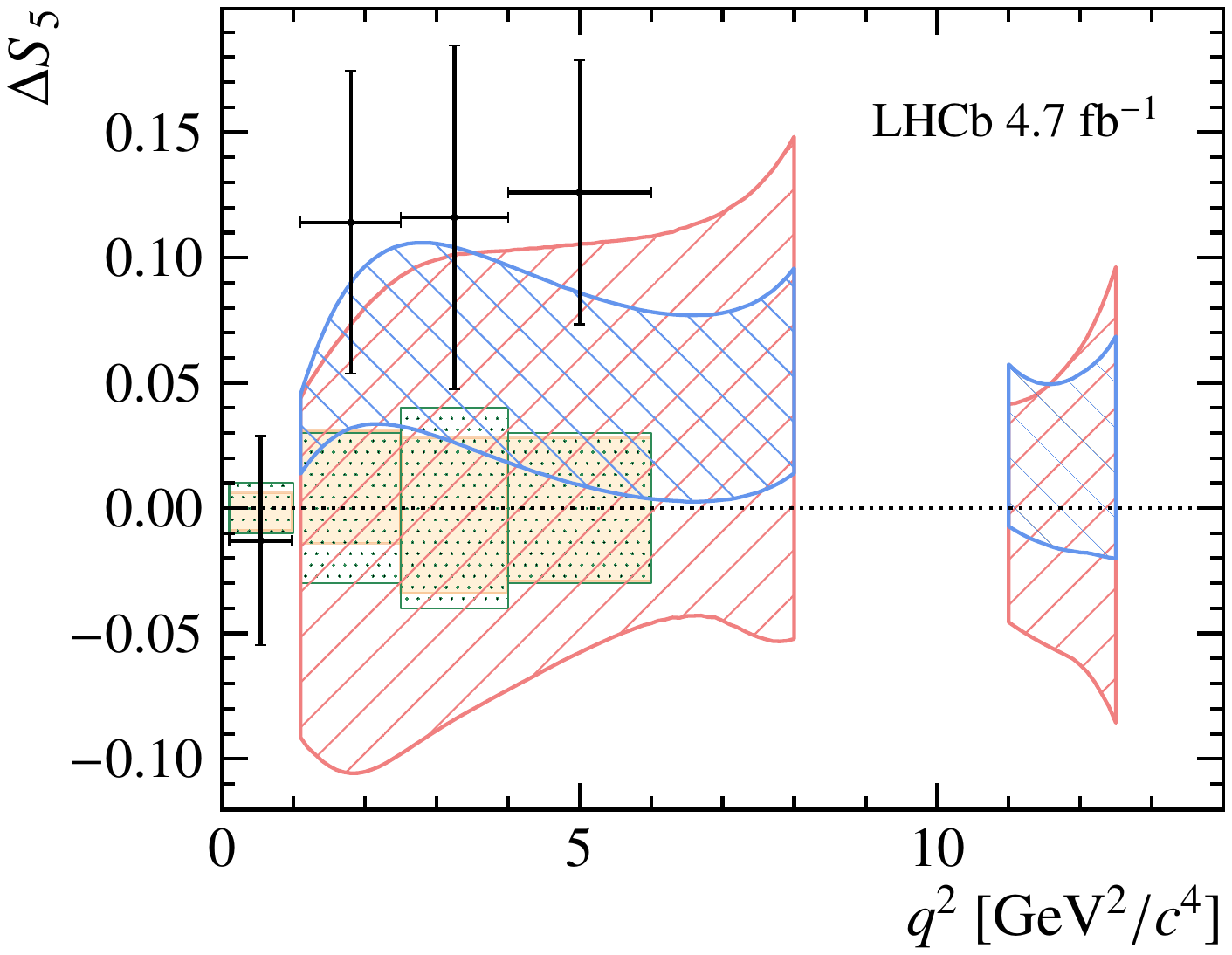}\\
    \vspace{-2.5mm}
    \includegraphics[width=0.45\textwidth]{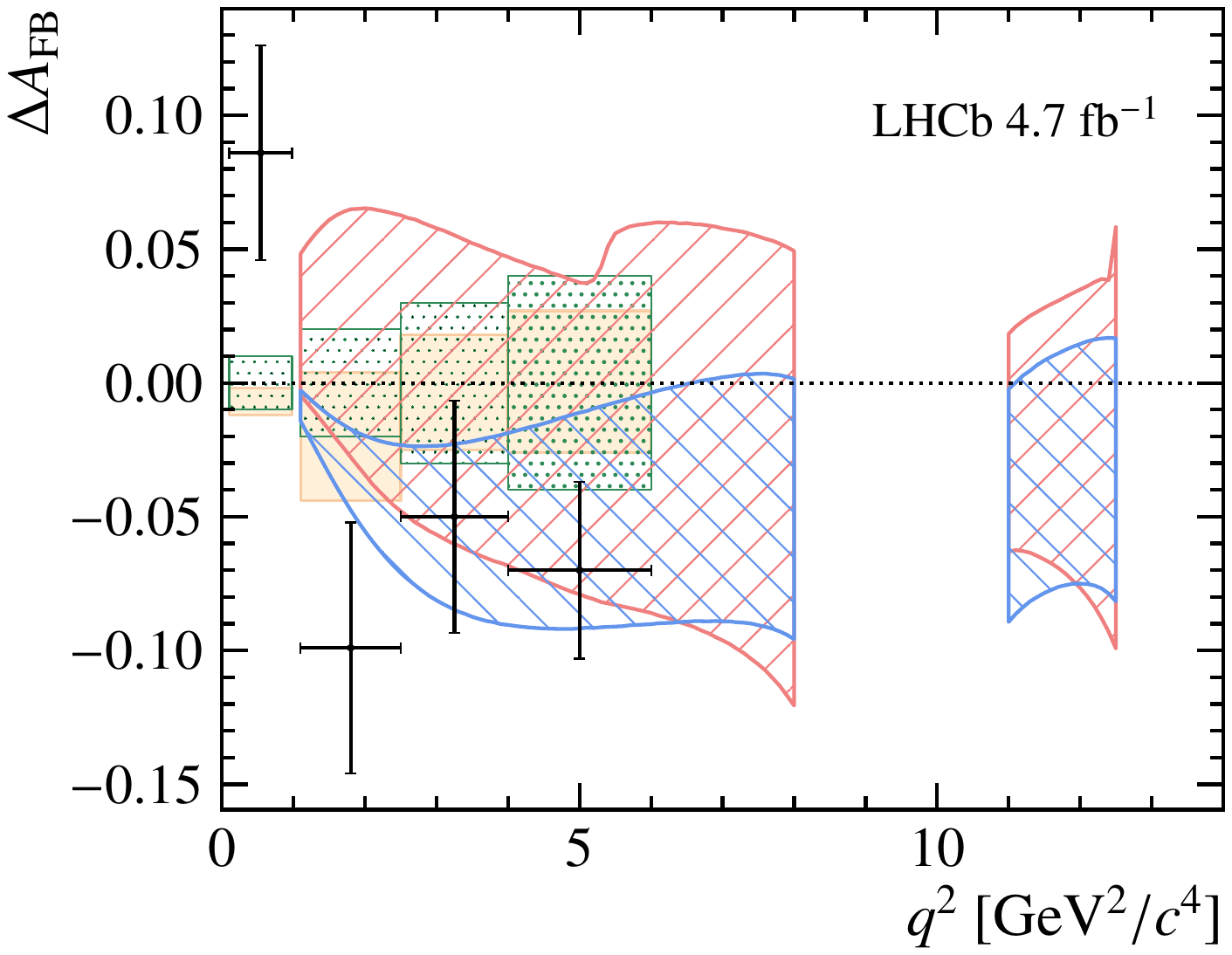}
    \hspace{-10mm}
    \includegraphics[width=0.45\textwidth]{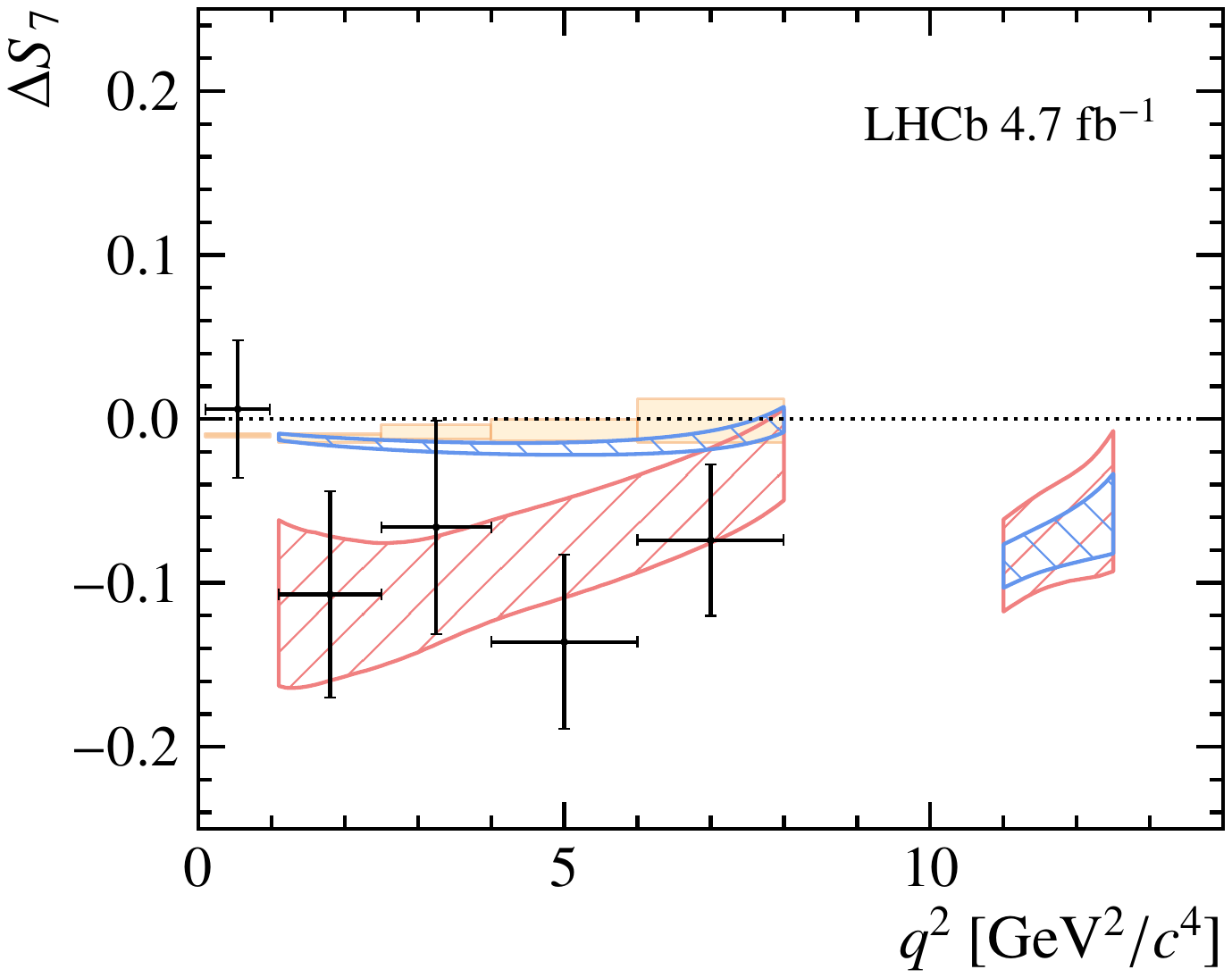}\\
    \vspace{-2.5mm}
    \includegraphics[width=0.45\textwidth]{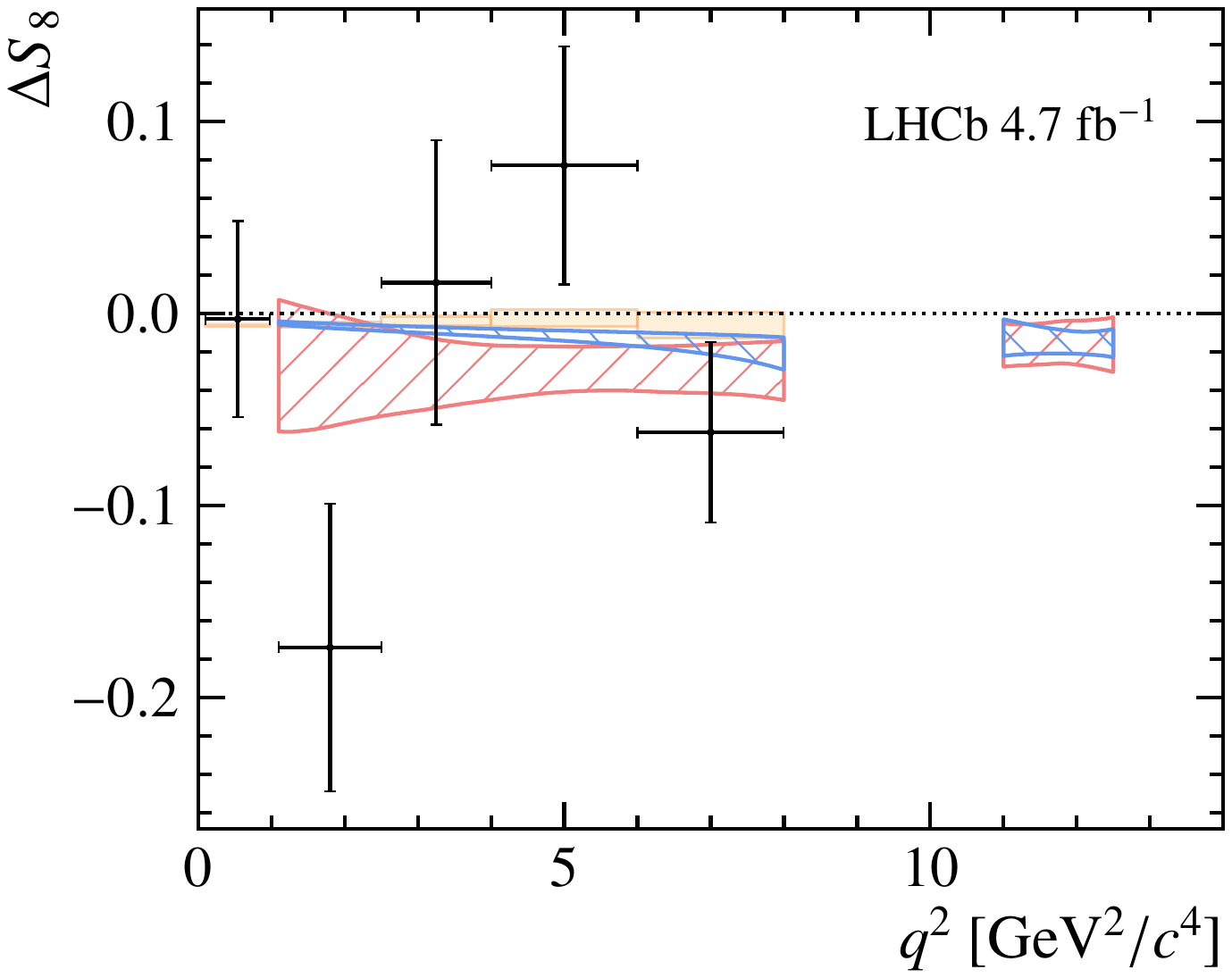}
    \hspace{-10mm}
    \includegraphics[width=0.45\textwidth]{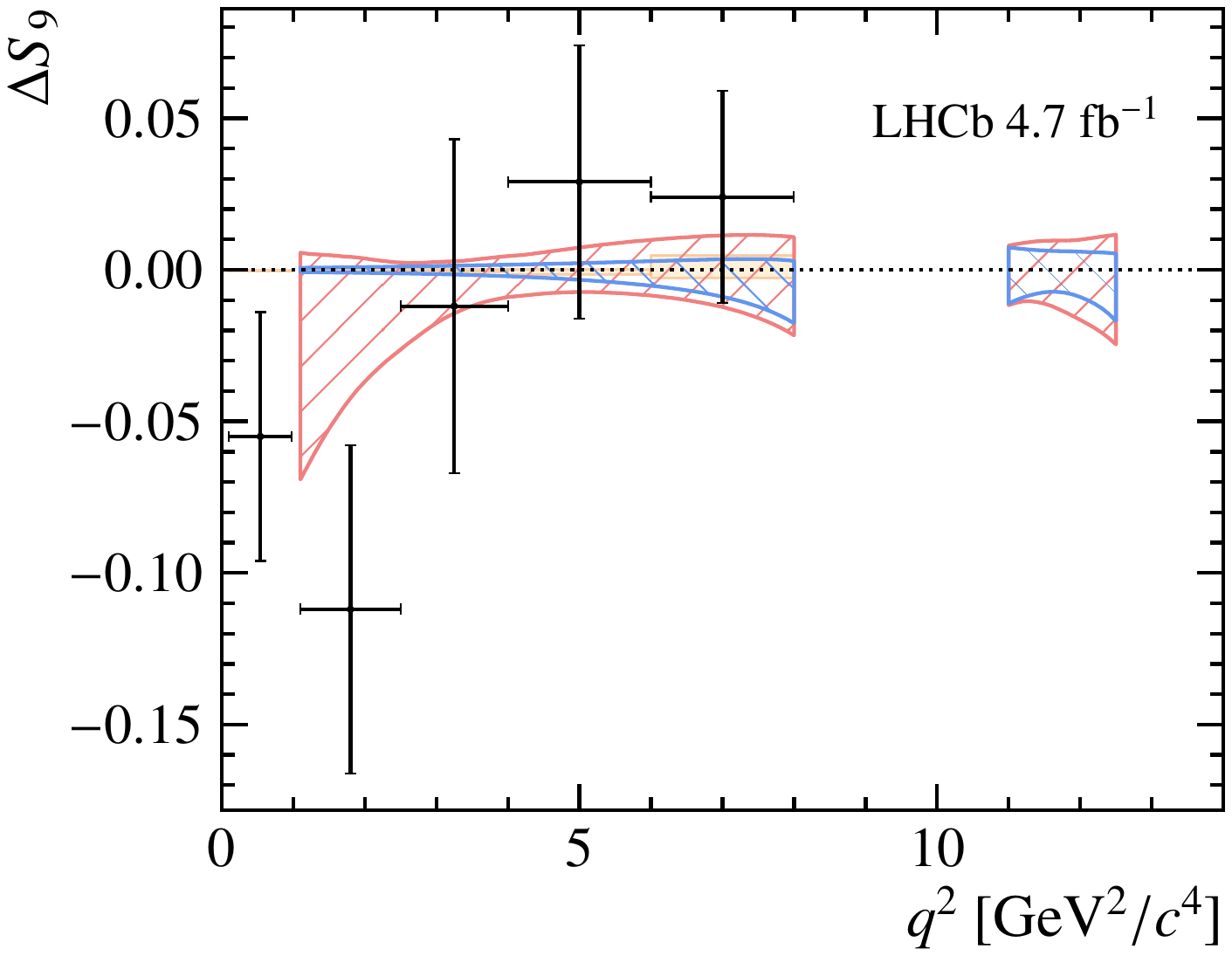}\\
\end{center}
  \vspace{-6mm}
  \caption{
  Non-local contributions to the angular observables (S-basis) obtained from data fits with (blue) and without (red) constraints from $q^{2} < 0$. The shaded regions are shown at 68\% confidence level. 
  When available, the difference between the LHCb binned angular analysis~\cite{LHCB-PAPER-2020-002}  
  and SM central-value from Refs.~\cite{Straub:2015ica,Altmannshofer:2014rta} is overlaid for reference (black dots), together with the uncertainty on the SM prediction (green) and difference between the SM predictions from Refs.~\cite{Gubernari:2018wyi,Gubernari:2022hxn,Horgan:2015vla} with respect to Refs.~\cite{Straub:2015ica,Altmannshofer:2014rta} (yellow). 
  }
  \label{fig:Sbasis}
\end{figure}

Figures~\ref{fig:Pbasis} and \ref{fig:Sbasis} show the impact of non-local hadronic contributions to the angular observables in the P- and S-basis, respectively.

\clearpage



\addcontentsline{toc}{section}{References}
\bibliographystyle{LHCb}
\bibliography{main,standard,LHCb-PAPER,LHCb-CONF,LHCb-DP,LHCb-TDR}

\newpage
\centerline
{\large\bf LHCb collaboration}
\begin
{flushleft}
\small
R.~Aaij$^{35}$\lhcborcid{0000-0003-0533-1952},
A.S.W.~Abdelmotteleb$^{54}$\lhcborcid{0000-0001-7905-0542},
C.~Abellan~Beteta$^{48}$,
F.~Abudin{\'e}n$^{54}$\lhcborcid{0000-0002-6737-3528},
T.~Ackernley$^{58}$\lhcborcid{0000-0002-5951-3498},
B.~Adeva$^{44}$\lhcborcid{0000-0001-9756-3712},
M.~Adinolfi$^{52}$\lhcborcid{0000-0002-1326-1264},
P.~Adlarson$^{78}$\lhcborcid{0000-0001-6280-3851},
C.~Agapopoulou$^{46}$\lhcborcid{0000-0002-2368-0147},
C.A.~Aidala$^{79}$\lhcborcid{0000-0001-9540-4988}
% ...

{\footnotesize \it

% ...
$^{35}$Nikhef National Institute for Subatomic Physics, Amsterdam, Netherlands\\
% ...
$^{44}$Instituto Galego de F{\'\i}sica de Altas Enerx{\'\i}as (IGFAE), Universidade de Santiago de Compostela, Santiago de Compostela, Spain\\
% ...
$^{46}$European Organization for Nuclear Research (CERN), Geneva, Switzerland\\
% ...
$^{48}$Physik-Institut, Universit{\"a}t Z{\"u}rich, Z{\"u}rich, Switzerland\\
% ...
$^{52}$H.H. Wills Physics Laboratory, University of Bristol, Bristol, United Kingdom\\
% ...
$^{54}$Department of Physics, University of Warwick, Coventry, United Kingdom\\
% ...
$^{57}$School of Physics and Astronomy, University of Glasgow, Glasgow, United Kingdom\\
$^{58}$Oliver Lodge Laboratory, University of Liverpool, Liverpool, United Kingdom\\
% ...
$^{66}$Syracuse University, Syracuse, NY, United States\\
% ...
$^{78}$Department of Physics and Astronomy, Uppsala University, Uppsala, Sweden, associated to $^{57}$\\
$^{79}$University of Michigan, Ann Arbor, MI, United States, associated to $^{66}$\\
% ...
}
\end{flushleft}

\end{document}